\documentclass[journal=nalefd,manuscript=article]{achemso}

\usepackage{graphicx}
\usepackage{dcolumn}
\usepackage{bm}
\usepackage{gensymb}
\usepackage{braket}
\usepackage{chemformula}
\usepackage[T1]{fontenc}

\usepackage{physics}
\usepackage{amsmath} 
\usepackage[colorlinks,allcolors=blue,urlcolor=blue]{hyperref}

\renewcommand{\qq}{\bm{q}}

\renewcommand{\vec}[1]{\bm{#1}}

\author{Dhruv C. Desai}
\altaffiliation{D.D. and J.P. contributed equally to this work}
\affiliation{Department of Applied Physics and Materials Science, California Institute of Technology, Pasadena, California 91125, USA.}

\author{Jinsoo Park}
\altaffiliation{D.D. and J.P. contributed equally to this work}
\affiliation{Department of Applied Physics and Materials Science, California Institute of Technology, Pasadena, California 91125, USA.}

\author{Jin-Jian Zhou}
\affiliation{School of Physics, Beijing Institute of Technology, Beijing 100081, China.}

\author{Marco Bernardi}
\email{bmarco@caltech.edu}
\affiliation{Department of Applied Physics and Materials Science, California Institute of Technology, Pasadena, California 91125, USA.}
\alsoaffiliation{Department of Physics, California Institute of Technology, Pasadena, California 91125, USA}

\title{Dominant two-dimensional electron-phonon interactions in the bulk Dirac semimetal Na$_3$Bi}

\begin{document}
\begin{abstract}
Bulk Dirac semimetals (DSMs) exhibit unconventional transport properties and phase transitions due to their peculiar low-energy band structure. 
Yet the electronic interactions governing nonequilibrium phenomena in DSMs are not fully understood. 
Here we show that electron-phonon ($e$-ph) interactions in a prototypical bulk DSM, Na$_3$Bi, are predominantly two-dimensional (2D). 
Our first-principles calculations discover a 2D optical phonon with strong $e$-ph interactions associated with in-plane vibrations of Na atoms. We show that this 2D mode governs $e$-ph scattering and charge transport in Na$_3$Bi, and induces a dynamical phase transition to a Weyl semimetal. 
\mbox{Our work} advances quantitative analysis of electron interactions in topological semimetals and reveals dominant low-dimensional interactions in bulk quantum materials. 
\end{abstract} 
\section{}
\begin{figure*}[t]
\includegraphics[width=1.0\textwidth]{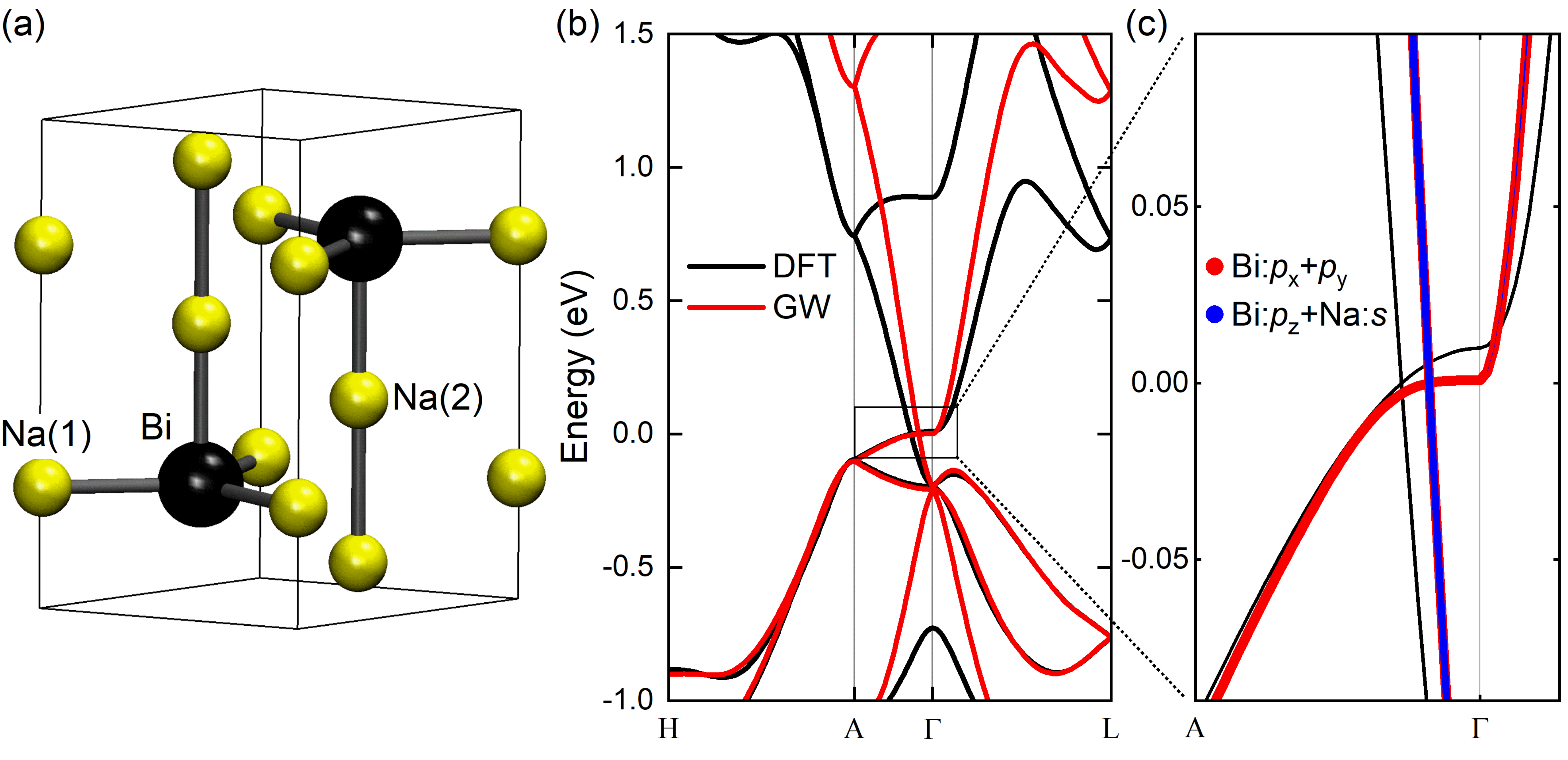}
\caption{
(a) Crystal structure of Na$_3$Bi with P$6_3/mmc$ space-group symmetry. The Bi (Na) atoms are shown with black (yellow) spheres. 
(b) Band structure of Na$_3$Bi comparing DFT (black) and GW (red) results. 
(c) Zoom-in of the band structure in (b) near the Dirac point, with states color-coded according to their orbital character. 
}\label{fig:bands}
\end{figure*}

\begin{figure}[!t]
\includegraphics[width=0.95\columnwidth]{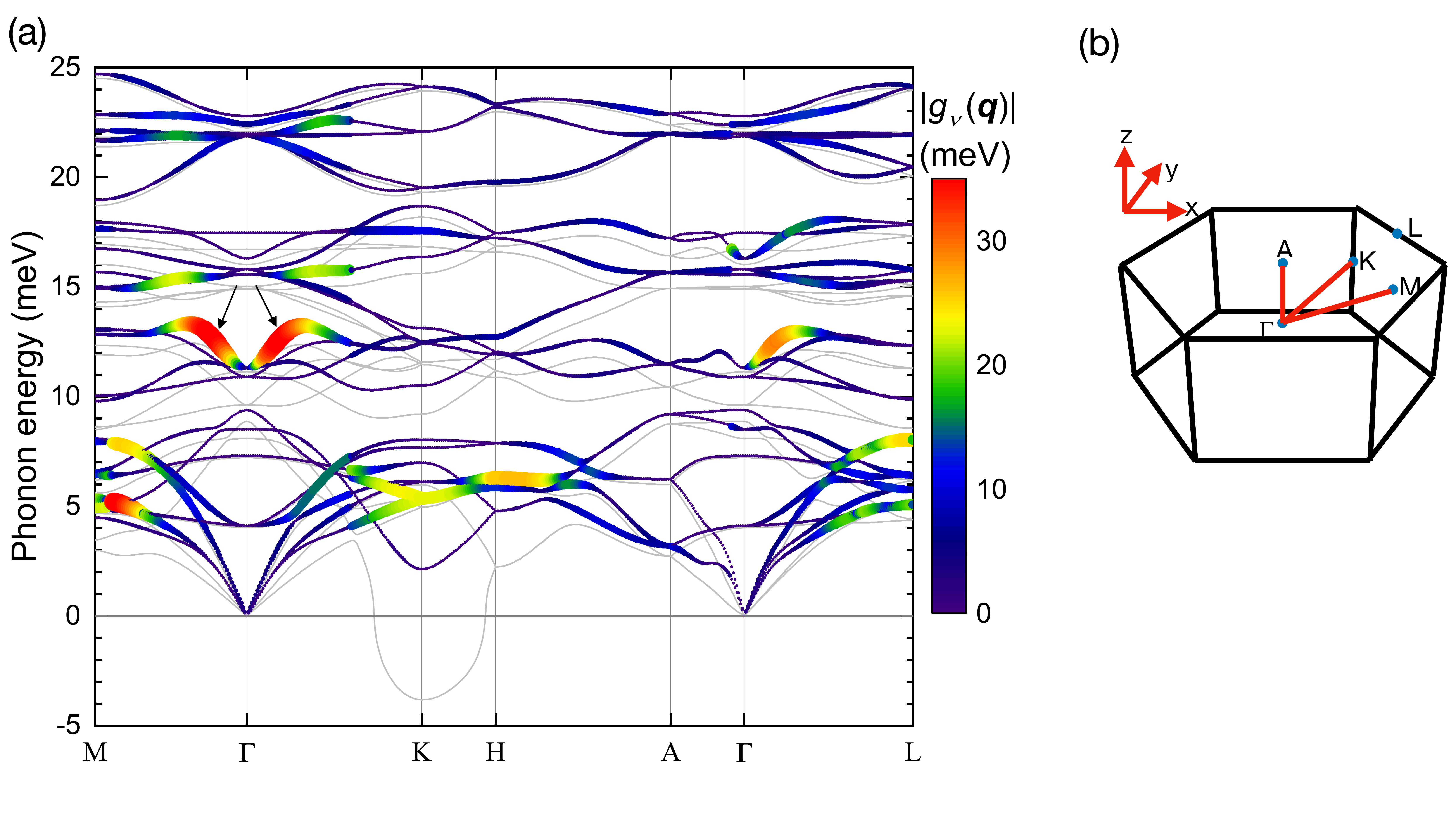}
\caption{(a) Na$_3$Bi phonon dispersion overlaid with a color map of the $e$-ph coupling strength, $\abs{g_\nu(\bm{q})}$, for wave-vector $\bm{q}$ along high-symmetry lines. 
The marker size is proportional to $\abs{g_\nu(\bm{q})}$ and the arrows indicate the strongly-coupled 2D phonon mode.
The phonon dispersion obtained with the simple acoustic sum rule is shown for comparison using gray lines, with imaginary frequencies shown as negative values. (b) Brillouin zone of Na$_3$Bi, shown to aid the interpretation of panel (a). The $\Gamma-M$ and $\Gamma-K$ directions are in the $xy$-plane, and $\Gamma-A$ corresponds to the $z$-direction. 
}\label{fig:phdisp}
\end{figure}

\begin{figure*}[!t]
\includegraphics[width=0.8\linewidth]{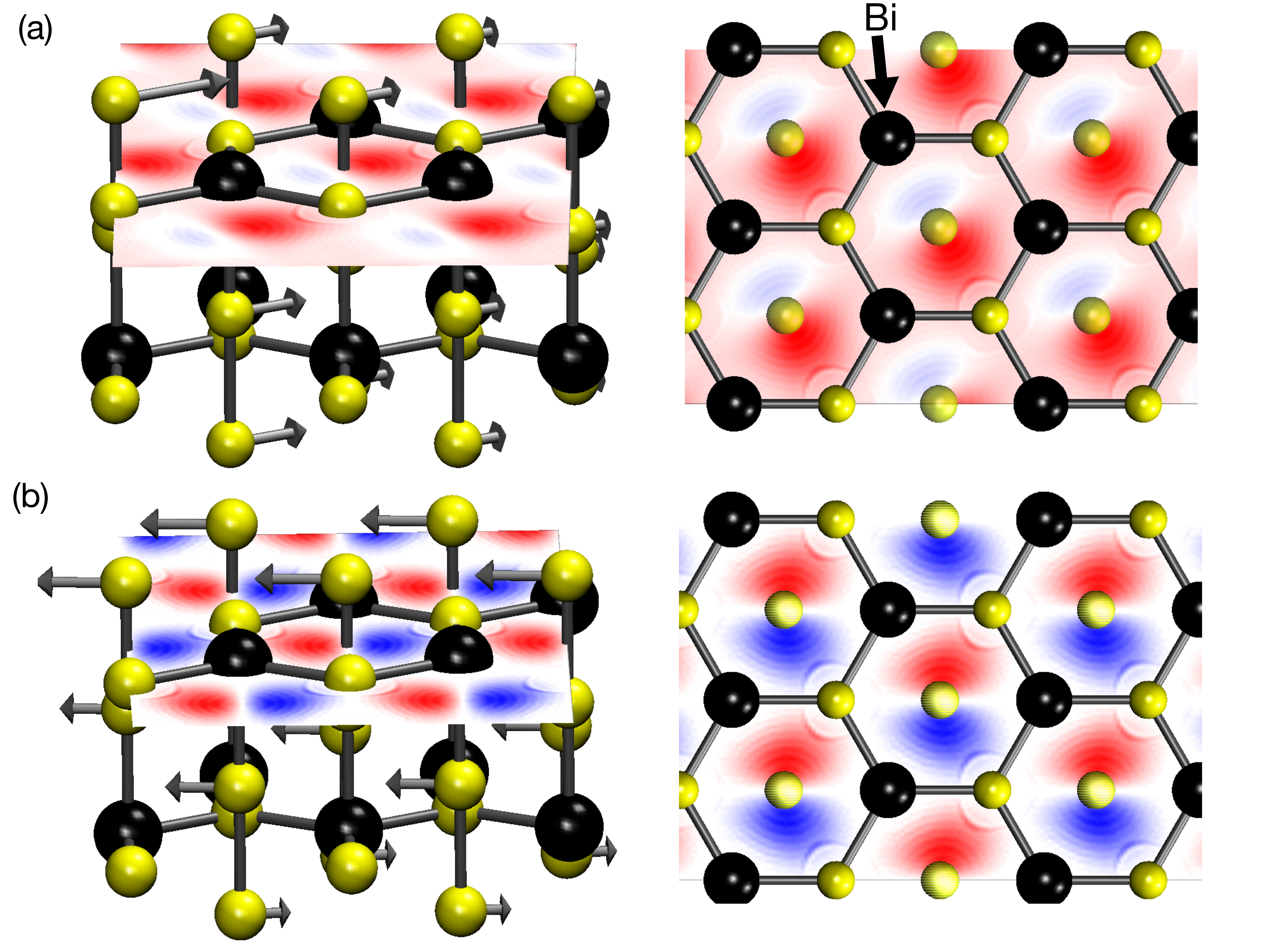}
\caption{Side view (left) and top view (right) of atomic motions and $e$-ph perturbation potentials $\Delta V_{\nu\bm{q}}(\mathbf{r})$ for two phonon modes: 
(a) Strongly-coupled 2D longitudinal optical phonon with wave-vector $\bm{q}$=(1/6,0,0) in the $xy$-plane, associated with a large perturbation at the Bi site leading to strong $e$-ph interactions; (b) transverse optical mode with $\bm{q}$ = (0,0,1/8) along the $z$-axis, resulting in a negligible $e$-ph coupling due to the weak perturbation at the Bi site. 
In both cases, we compute $\Delta V_{\nu\bm{q}}(\mathbf{r})$ from Na(2) atomic vibrations and plot it in the $xy$-plane containing Bi and Na(1) atoms. Red and blue colors correspond to positive and negative values of $\Delta V_{\nu\bm{q}}(\mathbf{r})$, respectively. 
}  
\label{fig:md12}
\end{figure*}

\indent
Topological semimetals are characterized by electronic band crossings near the Fermi energy, which result in linear band dispersions and  topologically nontrivial band structures~\cite{Armitage2018_WeylDirac_review}. There is a vast literature on their unusual properties, including high mobility and magnetoresistance~\cite{He_Cd3As2_2014,Liang_Cd3As2_2015,Cao_Cd3As2_2015,FengCd3As2_2015,Narayanan_Cd3As2_2015,Amarnath_Cd3As2_2019,Xiong_chiralNa3Bi_2015}, anomalous transport regimes~\cite{Xiong_chiralNa3Bi_2015,xiongAnomalous2016,Li2015_Cd3As2_chiralanomaly}, surface Fermi arcs~\cite{ZhijunWang2012_Na3Bi,LiangArpesNa3Bi_2016,Xu_Na3Bi_Fermiarcs2015,Moll2016_Cd3As2_fermiarcs}, and topological phase transitions~\cite{Collins2018_Na3bitransistor,Xia_Na3BiSTM_2019}. 
The discovery of graphene $-$ a two-dimensional Dirac semimetal (DSM) $-$ has enabled studies of new physics in a carbon atom sheet~\cite{geim2007rise}. 
In contrast with graphene, three-dimensional (bulk) DSMs are materials with rich structural and chemical complexity. They present a wide range of possible crystal structures and arrangements of Dirac cones, whose degeneracy is protected by crystal symmetry~\cite{Armitage2018_WeylDirac_review}, which makes bulk DSMs interesting for device applications~\cite{Dirac_appl_2017, spin_charge_2021}. 
\\
\indent 
Although many properties of DSMs can be explained using model low-energy Hamiltonians, the interactions between electrons and other degrees of freedom $-$ such as phonons, photons, and spin $-$ are not simple to quantify and give rise to rich physics in DSMs. \mbox{Examples} include phonon nonlinearities, unconventional nonequilibrium dynamics, and topological phase transitions,~\cite{Sie2019,Zhang_MoTe2_2019,Vaswani_2020,Xiao2020,Disa2021} among others. Electron-phonon ($e$-ph) interactions play a central role in this physics, but their understanding in bulk DSMs $-$ and more generally in topological semimetals $-$ is rather limited and relies mainly on phenomenological models~\cite{Yu_zrcosn_2021,Liu_biphenylene_2021}. 
First-principles calculations of $e$-ph interactions, which have now been applied to many classes of materials~\cite{Bernardi_2016,Agapito_2018,Bernardi_Si_2014,Zhou2016,Zhou_STO_2018,Zhou_STO_2019,Park_graphene_2007,Floris_Pb_2007,Li2015,Chen2017,Li2018, Giustino2018}, have been hindered in bulk DSMs by their complex atomic and electronic structures. 
\\
\indent
Sodium bismuthate ($\mathrm{Na_{3}Bi}$) is a prototypical bulk DSM  ~\cite{ZhijunWang2012_Na3Bi,liuDiscovery2014} 
whose Dirac cones have been observed by scanning tunneling spectroscopy~\cite{kushwahaBulk2015,Edmonds_Na3BiSTM_2017,Xia_Na3BiSTM_2019,dibernardoImportance2020}, angle-resolved photoemission (ARPES)~\cite{liuDiscovery2014,LiangArpesNa3Bi_2016}, and transport measurements~\cite{Xiong_chiralNa3Bi_2015}.
In Na$_3$Bi, first-principles calculations have examined impurity-limited transport~\cite{YuanQuantum2019}, nonequilibrium dynamics~\cite{Hubener2017}, and spin-orbit coupling~\cite{Tancogne-Dejean2022}.
However, a quantitative analysis of $e$-ph interactions is still missing. The interplay between crystal symmetry and electron spin, orbital, and momentum degrees of freedom suggests that $\mathrm{Na_{3}Bi}$ and other bulk DSMs may host unconventional $e$-ph interactions yet to be discovered. 
We explore this direction by carrying out a detailed first-principles study of $e$-ph interactions in Na$_3$Bi. 
We use density functional theory (DFT)~\cite{Martin-book,Baroni_DFPT_2001} to obtain the electronic structure, lattice  vibrations, and their interactions; our calculations take into account spin-orbit coupling (SOC) and many-body corrections to the electronic band structure (with the GW method~\cite{Hybertsen_GW_1986}), and employ an improved treatment of acoustic phonons (see Methods).
\\
\indent
Leveraging these accurate tools, we discover a dominant two-dimensional (2D) $e$-ph interaction in Na$_3$Bi associated with a 2D optical phonon with $e$-ph coupling strength far greater than that of any other mode.  
Our analysis shows that this 2D $e$-ph interaction governs the scattering and transport of Dirac electrons,  and reveals its microscopic origin. 
Similar \lq\lq killer'' phonon modes with dominant $e$-ph coupling controlling charge transport have been found in organic crystals~\cite{killer-ph} but not in topological materials. 
We also find that the strongly-coupled 2D mode breaks inversion symmetry in Na$_3$Bi and induces a dynamical phase transition to a Weyl semimetal (WSM). This finding points to new opportunities for ultrafast control of topological materials~\cite{Hubener2017,Sie2019,Zhang_MoTe2_2019,Vaswani_2020,Xiao2020,Disa2021}. 
\\
\indent
The unit cell of Na$_3$Bi, shown in Figure~\ref{fig:bands}a, belongs to the hexagonal $P6_{3}/mmc$ space group. Its crystal structure alternates a layer of Bi plus Na atoms, labeled Na(1), and a layer made up only by Na atoms, labeled Na(2). 
Inversion plus $C_3$ rotational symmetry result in a four-fold band degeneracy near the Fermi energy, with contributions from Na $3s$ and Bi $6p$ orbitals~\cite{ZhijunWang2012_Na3Bi,Jenkins_Na3Bi_2016,Shao_Na3Bibs_2017,Chiu_Na3Bi_2020,dibernardoImportance2020}. The Dirac cone is made up by two electronic bands, one with Na $3s$ $\!+\!$ Bi $6p_z$ and the other with Bi $6p_x$ $\!+\!$ $6p_y$ orbital character (the latter is denoted below as Bi-$p_{xy}$ band). 
To obtain an accurate band structure, we start from DFT and then apply a one-shot GW correction (see Methods), which 
increases the velocity of the Na $3s$ $\!+\!$ Bi $6p_z$ band by a factor of 1.8 and reduces the velocity of the Bi $p_{xy}$-band relative to DFT (Figure~\ref{fig:bands}b,c). 
The Fermi velocity computed with GW about 300~meV above the Dirac node is 7.0 $\!\times\!$ $10^5$~ms$^{-1}$, in excellent agreement with the experimental value of 8.1 $\!\times\!$ $10^5$~ms$^{-1}$~\cite{xiongAnomalous2016}. \mbox{Our computed} GW band structure agrees well with ARPES measurements by Liang et al.~\cite{LiangArpesNa3Bi_2016} (See Supplementary information). 
\\
\indent
The computed phonon dispersion in Na$_{3}$Bi is shown in Figure~\ref{fig:phdisp}a. 
The phonon frequencies are positive for all modes (color-coded curves in Figure~\ref{fig:phdisp}a), indicating a dynamically stable $P6_{3}/mmc$ crystal structure with no soft modes or imaginary frequencies. Fine-tuning the acoustic sum rule is crucial to obtaining this well-behaved phonon dispersion.  
Our results employ an advanced acoustic sum rule which minimally affects the inter-atomic force constants from DFPT~\cite{Mounet_asr_2005}; conversely,  
a widely-used $-$ so-called \lq\lq simple\rq\rq~$-$ acoustic sum rule~\cite{Giannozzi_QE_2009}, which modifies the inter-atomic force constants to enforce translational symmetry, leads to spurious soft phonons near the K-point of the Brillouin zone~\cite{chengGroundstate2014} (gray curves in Figure~\ref{fig:phdisp}a).
\\
\indent
\mbox{Our settings}, which combine a stable crystal structure, well-defined phonon dispersions, and electronic states with an accurate Fermi velocity allow us to carry out reliable first-principles calculations of $e$-ph interactions in Na$_3$Bi~\cite{Zhou_STO_2018}. 
We compute the $e$-ph matrix elements $g_{mn\nu}(\bm{k},\bm{q})$, which encode the $e$-ph coupling between pairs of electronic states (initial state $\ket{n\bm{k}}$ and final state $\ket{m\bm{k}+\bm{q}}$, where $n$ and $m$ are band indices and $\bm{k}$ the electron crystal momentum) due to a phonon with mode index $\nu$ and wave-vector $\bm{q}$. These calculations are carried out with the {\sc Perturbo} code~\cite{zhouPerturbo2020} as described in Methods. 
\\
\indent
Figure~\ref{fig:phdisp} shows the phonon dispersion in Na$_{3}$Bi overlaid with a color map of the $e$-ph coupling strength, 
defined as $\abs{g_\nu(\bm{q})}$ $\equiv$ $(\sum_{mn}\abs{g_{mn\nu}(\bm{k}=0,\bm{q})}^2/N_{b})^{1/2}$ (here we sum over $N_{b}=2$ lowest conduction bands)~\cite{zhouPerturbo2020}.  
We find that the $e$-ph interactions are overall relatively weak in Na$_{3}$Bi, with an average value $\abs{g_\nu(\bm{q})} \approx$ 5~meV. 
\mbox{Yet one particular} phonon mode, with $\sim$12~meV energy and wave-vector $\bm{q}$ in the $\Gamma$$-$M and $\Gamma$$-$K directions (which correspond to the crystal $xy$-plane; see Figure~\ref{fig:phdisp}b) exhibits a much stronger $e$-ph coupling than any other mode, with value $\abs{g} \approx 35~\text{meV}$. 
This strongly-coupled 2D mode is a longitudinal optical (LO) phonon that is infrared-active and has $E_{1u}$ character at the zone center~\cite{Pizzi_shearbreathing_2021}. Its associated atomic vibrations, shown in Figure~\ref{fig:md12}a, have primary contributions from Na(2) atoms, which oscillate with large amplitudes in the Na-only layers of Na$_3$Bi, and have negligible contributions from the Na(1) and Bi atoms in the neighboring layers. 
Because the wave-vector and atomic displacements of this strongly-coupled 2D mode are both in the $xy$-plane, the dominant $e$-ph interactions in Na$_3$Bi are inherently two-dimensional. 
\\
\indent
To understand the microscopic origin these strong 2D $e$-ph interactions, we analyze their perturbation potential, whose local lattice-periodic part can be written as~\cite{Baroni_DFPT_2001,zhouPerturbo2020}  
\begin{equation}\label{perturbation_pot}
   \Delta V_{\nu\bm{q}}(\mathbf{r}) \equiv  \sum_{\kappa}\frac{1}{\sqrt{M_{\kappa}}}\, \mathbf{e}^{(\kappa)}_{\nu\bm{q}} \cdot \partial_{\kappa, \bm{q}} V(\mathbf{r})\,\,, \vspace{-6pt}
\end{equation}
where $M_{\kappa}$ is the mass and $\mathbf{e}^{(\kappa)}_{\nu\bm{q}}$ the displacement eigenvector of atom $\kappa$ due to phonon mode $(\nu,\bm{q})$, and $\partial_{\kappa, \bm{q}} V(\mathbf{r})$ is the derivative of the local Kohn-Sham potential with respect to the position of atom $\kappa$~\cite{zhouPerturbo2020}. 
We focus on the effect of the dominant Na(2) atomic vibrations on the Bi-$p_{xy}$ Dirac-cone electronic states near $\Gamma$. 
Figure~\ref{fig:md12} shows the $e$-ph perturbation potential $\Delta V_{\nu\bm{q}}(\mathbf{r})$ generated by Na(2) atomic vibrations and plotted in the $xy$-plane containing Bi atoms.
For the dominant 2D LO mode (Figure~\ref{fig:md12}a), which has wave-vector $\bm{q}$ in the $xy$-plane, the Na(2) atoms move out-of-phase within each layer, causing large perturbations at the Bi atoms. As a result electronic states in the Bi $p_{xy}$-band couple strongly with this phonon mode.
Increasing $\abs{\bm{q}}$ in the $xy$-plane leads to an even greater perturbation at the Bi site and thus stronger \mbox{$e$-ph coupling.} 
\mbox{In contrast, for a 2D transverse optical} mode propagating in the $z$-direction the Na(2) atoms move uniformly in-phase in the $xy$-plane (Figure~\ref{fig:md12}b).  
In this case, $\Delta V_{\nu\bm{q}}(\mathbf{r})$ has a symmetric pattern with nodes at Bi atoms, which suppresses $e$-ph coupling for the Bi-$p_{xy}$ band. 
Accordingly, we find a very weak $e$-ph coupling for such transverse optical modes, as shown by the dark blue color in the $\Gamma$$-$A direction in Figure~\ref{fig:phdisp}a. 
\\
\indent
We analyze two important consequences of the strong 2D $e$-ph coupling in Na$_3$Bi. First, we find that charge transport is governed by scattering of Dirac electrons with the strongly-coupled 2D mode, which contributes nearly half of the total $e$-ph scattering rate (Figure~\ref{fig:trans}a) and resistivity (see below). 
Other individual phonon modes contribute significantly less, up to 15\% of the $e$-ph scattering rate for the mode with the second strongest coupling. Therefore, this strongly-coupled 2D LO mode is analogous to the \lq\lq killer'' phonons controlling charge transport recently discovered in organic crystals~\cite{killer-ph}. 
\\

\indent
Because of the strong 2D $e$-ph coupling, Dirac-cone electronic states with in-plane momentum $\bm{k}$, which couple to each other via phonons with in-plane momenta, exhibit large $e$-ph scattering rates (Figure~\ref{fig:trans}b). In contrast, electrons with momentum $\bm{k}$ in the $z$-direction scatter mostly via phonons with out-of-plane $\bm{q}$, and are associated with smaller scattering rates. This anisotropic scattering due to 2D $e$-ph coupling is evident in the entire temperature range we analyzed (77$-$300~K). We have verified that the $e$-ph matrix elements $g(\bm{k},\bm{q})$ possess a similar anisotropy, such that the $e$-ph coupling strength $|g(\bm{k},\bm{q})|$ is much greater for in-plane than for out-of-plane electron momenta.
\\
\indent
We compute the phonon-limited mobility and resistivity using these first-principles $e$-ph scattering rates in the Boltzmann transport equation~\cite{zhouPerturbo2020} (see Methods). 
Our results show that the in-plane mobility for temperatures between 150$-$400~K is very large (Figure~\ref{fig:trans}c) $-$ up to $\sim$30,000~$\text{cm}^2/\text{Vs}$ at room temperature and high electron concentration, mainly as a result of the high Fermi velocity of the Bi $p_{xy}$-band and the overall weak $e$-ph coupling. This mobility limit, which applies to an ideally pure crystal of Na$_3$Bi where charge transport is impeded only by phonons, is exceptionally high and has the same order of magnitude as the mobility in graphene~\cite{geim2007rise}. To our knowledge, such large electron mobilities have not yet been measured in Na$_3$Bi near room temperature; one possible reason is that Na$_3$Bi samples  typically contain large concentrations of defects, particularly Na vacancies, which may make the intrinsic phonon-limited mobility difficult to observe~\cite{kushwahaBulk2015,Edmonds_Na3BiSTM_2017}. Improvements in growth techniques may bring the experimental mobility of Na$_3$Bi closer to our predicted theoretical limit. Note that in Cd$_3$As$_2$, a widely studied DSM, mobility values as high as $\sim$40,000~$\text{cm}^2/\text{Vs}$ at 130~K have been reported~\cite{Neupane_Cd3As2_2014}, which are comparable to the $\sim$100,000~$\text{cm}^2/\text{Vs}$ we predict in Na$_3$Bi for the same temperature and carrier concentration. 
\\
\indent
To complete our discussion on transport, Figure~\ref{fig:trans}d shows the computed in-plane resistivity as a function of temperature for Fermi energies between 100$-$300~meV. In this regime, the transport behavior is metallic, and the resistivity increases with temperature following a power law. 
Comparison with experiments is important despite the variability in Na$_3$Bi sample quality noted above.
We compare our calculations with the measurements by Xiong \textit{et al}.~\cite{xiongAnomalous2016}, which achieve the lowest resistivity among available experimental data~\cite{xiongAnomalous2016,kushwahaBulk2015} indicating higher sample quality.
Our computed resistivity is lower than their measured values~\cite{xiongAnomalous2016} by about an order of magnitude at 50~K and a factor of 3$-$5 at 250~K. The lower discrepancy at higher temperature indicates an improved agreement between theory and experiment in the intrinsic, phonon-limited transport regime studied in this work. 
\begin{figure}[t!]
\includegraphics[width=0.95\columnwidth]{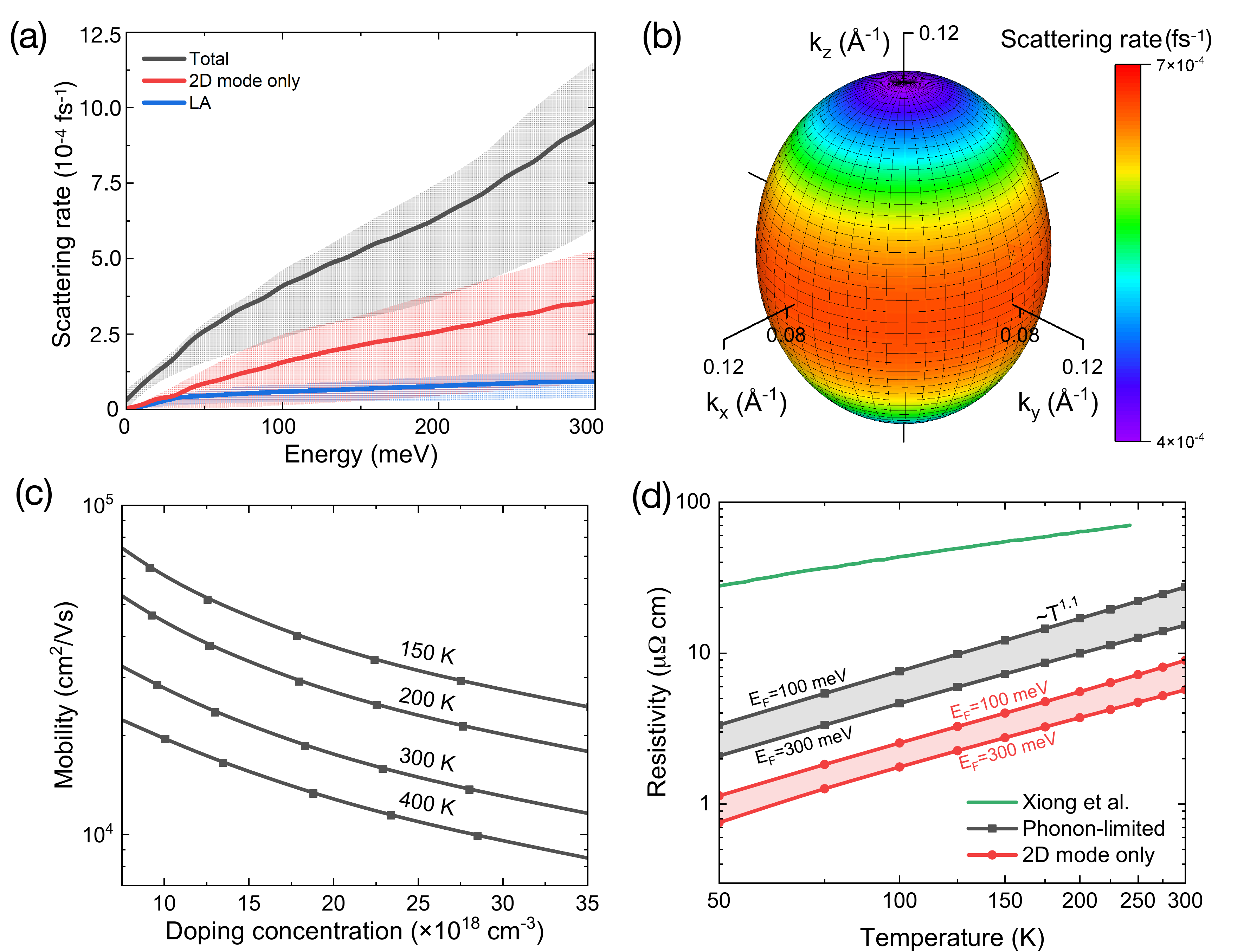}
\caption{Calculations of transport and $e$-ph scattering in Na$_3$Bi. (a) E-ph scattering rates as a function of carrier energy at 300~K and $E_{\rm F}$=200~meV. We show the total scattering rate (black) and the contributions from the strongly-coupled 2D mode (red) and the longitudinal acoustic (LA) mode with second strongest coupling (blue). For each curve, we plot the average scattering rate with a solid line, and show the standard deviation as a shaded region. (b) Fermi surface at $E_{\rm F}$=200~meV color-coded according to the total $e$-ph scattering rates. (c) Electron mobility as a function of carrier concentration for temperatures between 150$-$400~K. 
(d) Temperature dependent resistivity for Fermi energies between 100$-$300~meV above the Dirac point, shown together with the strongly-coupled 2D-mode contribution. Experimental results by Xiong et al.~\cite{xiongAnomalous2016} are shown for comparison. 
}\label{fig:trans}
\end{figure}
\\
\indent 
It is interesting to compare these findings with graphene, a 2D DSM. In both Na$_3$Bi and graphene, a 2D optical phonon has the strongest $e$-ph coupling~\cite{Park_graphene_2014,Xiao_graphene_2021}. However, heavier atoms and weaker bonding in Na$_3$Bi result in softer phonons $-$ the energy of the strongly-coupled 2D phonon in Na$_3$Bi is only $\sim$12~meV, and thus much smaller than the $\sim$200 meV energy of strongly-coupled 2D optical phonons in graphene~\cite{Xiao_graphene_2021}. 
\mbox{At room temperature,} where $k_{\rm B}T\!\approx\!26$~meV, the strongly-coupled 2D phonons are thermally excited in Na$_3$Bi, while in graphene only acoustic phonons are present. As a result, optical modes contribute less than 15\% to the resistivity in graphene at 300~K~\cite{desaiMagnetotransport2021}, versus a dominant 50\% resistivity contribution from the strongly-coupled 2D phonon in Na$_3$Bi (Figure~\ref{fig:trans}d). 
Note that while graphene is a 2D material, Na$_3$Bi is a bulk crystal where a dominant 2D $e$-ph interaction is unexpected. 
\begin{figure}[t!]
\includegraphics[width=0.95\columnwidth]{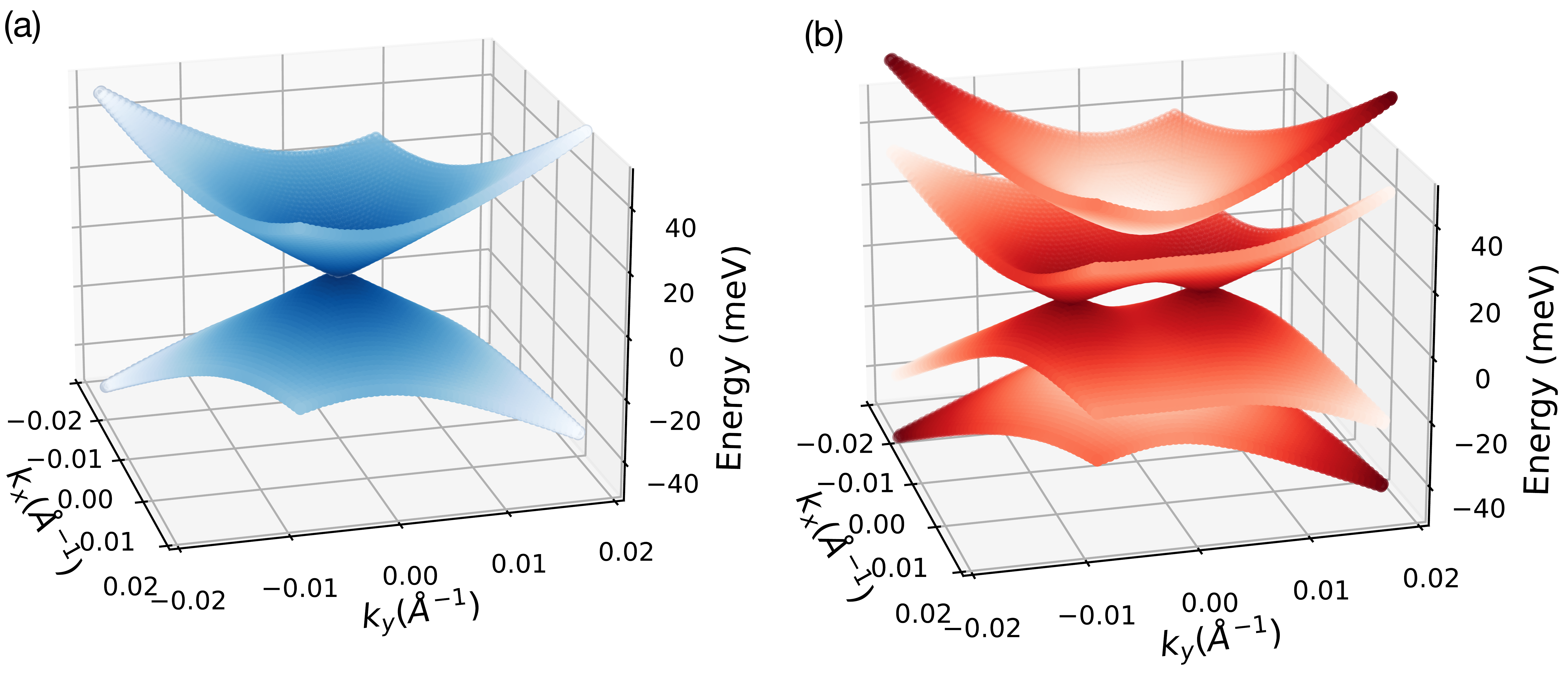}
\caption{Electronic band structure computed with DFT for (a) pristine and (b) 2D-mode distorted Na$_3$Bi. The displacement in (b) is 3\% of the in-plane lattice constant.  
}\label{fig:ti}
\end{figure}
\\
\indent
Finally, we find that the atomic vibrations from the strongly-coupled 2D mode dynamically induce a phase transition to a Weyl semimetal (WSM) in Na$_3$Bi. Due to its $E_{1u}$ character~\cite{Pizzi_shearbreathing_2021}, this 2D mode dynamically breaks inversion symmetry and removes the four-fold degeneracy at the Dirac point, splitting each Dirac cone into a pair of Weyl cones. 
Figure~\ref{fig:ti} shows the DFT band structure in the  $k_x$-$k_y$ plane containing the Dirac node, comparing results for the pristine structure (Figure~\ref{fig:ti}a) and for the lattice distorted from the strongly-coupled 2D mode (Figure~\ref{fig:ti}b), which is computed with frozen-in atomic displacements along the 2D-mode eigenvector (see Methods). 
The atomic displacements split each Dirac node into a pair of Weyl nodes separated along $k_y$ by about 0.01 $\text{\AA}^{-1}$; the system remains metallic throughout this phase transition. 
Inversion symmetry breaking from the strongly-coupled $E_{1u}$ mode is crucial to obtaining the WSM phase: phonon distortions that preserve inversion symmetry but break the threefold rotational symmetry $-$ for example, $E_{2g}$ modes, which are however weakly coupled $-$ are not robust against a gap opening, and instead cause a phase transition to a topological insulator.
\\
\indent
These results imply that 2D phonons with strong $e$-ph coupling can provide a versatile knob for ultrafast control in Na$_3$Bi. In particular, because the strongly-coupled 2D LO mode is infrared active~\cite{Dong_Na3BiPhonons_2019} (but not Raman active), one could induce a topological phase transition in Na$_3$Bi by coherently driving this 2D mode using a THz pulse~\cite{Merlin} or through carrier optical excitation followed by strong $e$-ph coupling~\cite{Merlin}.
Interestingly, Hübener et al.~\cite{Hubener2017} have shown that a similar phase transition from a DSM to a WSM can be achieved in Na$_3$Bi with a different mechanism $-$ strong light-matter coupling, which dresses the electronic states inducing Floquet-Weyl nodes.
These predictions contribute to the thriving area of driven nonequilibrium dynamics in topological materials, where recent experiments on bulk DSMs~\cite{Vaswani_2020} and WSMs~\cite{Sie2019, Zhang_MoTe2_2019} have demonstrated nonequilibium topological phase transition using electric fields or optical pulses.
\\
\indent
In conclusion, we have shown that the dominant $e$-ph interactions in a prototypical bulk DSM, Na$_3$Bi, are inherently two-dimensional and govern the scattering and transport of Dirac electrons. Our first-principles analysis reveals the microscopic origin of this strong 2D $e$-ph coupling; it also shows that the strongly-coupled 2D mode can induce a dynamical phase transition to a WSM, suggesting new routes for ultrafast control of Dirac electrons in bulk DSMs. 
These results seed the question of whether other bulk materials may host dominant low-dimensional $e$-ph interactions governing their physical properties. For example, MgB$_2$, a superconductor with a relatively high critical temperature of $\sim$40~K, has a crystal structure similar to Na$_3$Bi and has been hypothesized to host a 2D phonon with strong $e$-ph coupling~\cite{Bohnen_mgb2001,An_MgB2_2001}.
First-principles calculations such as those shown in this work can contribute to address these questions and advance future discoveries of electronic interactions and nonequilibrium dynamics in topological materials.
\newpage
\section*{Methods}\label{methods}
\indent
\subsection{DFT, DFPT and GW calculations}
We perform DFT calculations in a plane-wave basis set using the \textsc{Quantum ESPRESSO}~\cite{Giannozzi_QE_2009} code. 
We employ the PBEsol~\cite{pbesol2008} exchange-correlation functional and fully-relativistic norm-conserving pseudopotentials from Pseudo Dojo~\cite{pseudodojo}. These calculations use a coarse $12 \times 12 \times 8$ $\bm{k}$-point grid, a kinetic energy cutoff of 90~Ry, and relaxed lattice constants ($a = 5.42~\text{\AA}$ and $c = 9.67~\text{\AA}$)  which are in excellent agreement with the experimental values ($a = 5.45~\text{\AA}$ and $c = 9.66~\text{\AA}$)~\cite{ZhijunWang2012_Na3Bi}. The phonon dispersions and perturbation potentials are computed using coarse grids with $6 \times 6 \times 8$ $\bm{q}$-points using DFPT~\cite{Baroni_DFPT_2001}. We employ the crystal acoustic sum rule from Ref.~\cite{Mounet_asr_2005} to compute phonon dispersions. 
The GW correction to the electronic band structure is computed using the \textsc{yambo} code.~\cite{sangalliManybody2019} We employ 120 unoccupied bands and a 10~Ry energy cutoff for the dielectric screening combined with the Bruneval-Gonze terminator~\cite{Bruneval_GW_2008}; 
we have verified that increasing the number of unoccupied bands to 600 and the energy cutoff to 30~Ry has a negligible effect. 
\\
\indent
\subsection{Electron-phonon matrix elements and perturbation potentials}
 We use the \textsc{Perturbo} code~\cite{zhouPerturbo2020} to obtain the $e$-ph coupling matrix elements on the coarse $\bm{k}$- and $\bm{q}$-point grids given above. The $e$-ph matrix elements $g_{mn\nu}(\bm{k},\bm{q})$ are defined as 
\begin{equation}\label{eq:gdef}
g_{mn\nu}(\bm{k},\bm{q}) = \sqrt{\frac{\hbar}{2\omega_{\nu\bm{q}}}}\bra{\psi_{m\bm{k+q}}}\Delta V_{\nu\bm{q}}\ket{\psi_{n\bm{k}}},
\end{equation}
where $\ket{\psi_{n\bm{k}}}$ and $\ket{\psi_{m\bm{k+q}}}$ are Bloch states with momenta $\bm{k}$ and $\bm{k+q}$, and $\Delta\mathrm{V}_{\nu\bm{q}}$ is the lattice-periodic part of the phonon perturbation potential~\cite{zhouPerturbo2020}. Since the dominant contribution to $g(\bm{k},\bm{q})$ for the 2D mode comes from Na(2) atomic vibrations, we analyze the effect of Na(2) motions on the Bi $p_{xy}$ band, which is achieved by setting $\mathbf{e}^{(\kappa)}_{\nu\bm{q}}$ to 0 in eq~\ref{perturbation_pot} for Na(1) and Bi atoms. A similar analysis can be performed for Na(1) and Bi atomic motions, or using the Na $s$ + Bi $p_z$ band, but their contributions to $g(\bm{k},\bm{q})$ are significantly smaller and do not affect our conclusions.

\subsection{Electron-phonon scattering rates and charge transport}
We interpolate the $e$-ph matrix elements on fine BZ grids with up to $130 \times 130 \times 90$ $\bm{k}$- and $\bm{q}$-points using maximally localized Wannier functions~\cite{MarzariMLWF2012} generated with the Wannier90 code.~\cite{Mostofi_2014} 
We then compute the $e$-ph scattering rates at temperature $T$ using~\cite{zhouPerturbo2020} 
\begin{equation}\label{eq:scatrate}
\begin{split}
\Gamma_{n\bm{k}}(T)=&\frac{2\pi}{\hbar}\sum_{m \nu \qq}\abs{g_{mn\nu}(\bm{k},\qq)}^2\\
&[(N_{\nu \qq}+1-f_{m\vec{k+q}})\delta(\varepsilon_{n\bm{k}}-\varepsilon_{m\vec{k+q}}-\hbar\omega_{\nu\qq}) \\
&~+ (N_{\nu \qq}+f_{m\vec{k+q}})\delta(\varepsilon_{n\bm{k}}-\varepsilon_{m\vec{k+q}} +\hbar\omega_{\nu\qq})],
\end{split}
\end{equation}
where  $\varepsilon_{n\bm{k}}$ and $f_{m\vec{k+q}}$ are electron band energies and occupations, respectively, while $\hbar\omega_{\nu \bm{q}}$ and $N_{\nu \bm{q}}$ denote phonon energies and occupations. The electron and phonon  occupations govern the temperature dependence of the $e$-ph scattering rates.

Using \textsc{Perturbo}~\cite{zhouPerturbo2020}, we obtain the carrier mobility as a function of temperature and doping concentration by solving the linearized Boltzmann transport equation:~\cite{zhouPerturbo2020}
\begin{equation} \label{eq:Linearized_BTE}
      \bm{F}_{n\bm{k}} = \bm{v}_{n\bm{k}} \tau_{n\bm{k}} + \frac{\tau_{n\bm{k}}}{\mathcal{N}_{\bm{q}}} \sum_{m,\nu \bm{q}} W_{n\bm{k},m\bm{k+q}}^{\nu \bm{q}} \bm{F}_{m \bm{k+q}}
\end{equation}
where $W_{n\bm{k},m\bm{k+q}}^{\nu \bm{q}}$ are phonon mode-dependent $e$-ph scattering rates from electronic state $\ket{n\bm{k}}$ to $\ket{m\bm{k+q}}$ due to phonon mode $(\nu,\bm{q})$;  $\bm{v}_{n\bm{k}}$ are band velocities, $\tau_{n\bm{k}}$  are relaxation times, and $\mathcal{N}_{\bm{q}}$ is the number of $\bm{q}$-points used in the Brillouin zone summation. Above, $\bm{F}_{n\bm{k}}(T)$ is a term proportional to the first-order deviation of the electron occupations $f_{n\bm{k}}$ from their equilibrium values $f_{n\bm{k}}^0$ due to the electric field $\bm{E}$, and is defined through 
\begin{equation}
-f_{n\bm{k}}^0(1 - f_{n\bm{k}}^0)\frac{e\bm{E}}{k_{B}T} \cdot \bm{F}_{n\bm{k}} = f_{n\bm{k}} - f_{n\bm{k}}^0\vspace{-10pt}. 
\end{equation} The conductivity tensor $\sigma_{\alpha\beta}$ is computed using
\begin{equation}
\sigma_{\alpha\beta} = e^2 \int {dE\, (-\partial f^0 / \partial E)\, \Sigma_{\alpha\beta}(E,T)},
\end{equation}
where $\alpha$ and $\beta$ are Cartesian directions, and $\Sigma_{\alpha\beta}(E,T)$ is the transport distribution function at energy $E$ and temperature $T$:~\cite{Zhou2016}
\begin{equation}
    \Sigma_{\alpha\beta}(E,T)=\frac{1}{ \mathcal{N}_{\bm{k}} \Omega} \sum_{n\bm{k}}  \mathrm{v}_{n\bm{k}}^\alpha \bm{F}_{n\bm{k}}^{\beta}(T) \delta(E-\varepsilon_{n\bm{k}}).
\end{equation}
Here, $\mathcal{N}_{\bm{k}}$ is the number of $\bm{k}$-points in the Brillouin zone and $\Omega$ is the unit cell volume.  

\subsection{Acknowledgments}
\vspace{-10pt}
D.D. and J.P. thank Ivan Maliyov for fruitful discussions. This work was supported by the National Science Foundation under Grant No. DMR-1750613, which provided for method development, and Grant No. OAC-2209262, which provided for code development. M. B. was partially supported by the AFOSR and Clarkson Aerospace under Grant No. FA95502110460.
This research used resources of the National Energy Research Scientific Computing Center (NERSC), a U.S. Department of Energy Office of Science User Facility located at Lawrence Berkeley National Laboratory, operated under Contract No. DE-AC02-05CH11231.\\
\newpage
\providecommand{\latin}[1]{#1}
\makeatletter
\providecommand{\doi}
  {\begingroup\let\do\@makeother\dospecials
  \catcode`\{=1 \catcode`\}=2 \doi@aux}
\providecommand{\doi@aux}[1]{\endgroup\texttt{#1}}
\makeatother
\providecommand*\mcitethebibliography{\thebibliography}
\csname @ifundefined\endcsname{endmcitethebibliography}
  {\let\endmcitethebibliography\endthebibliography}{}


\begin{mcitethebibliography}{72}
\providecommand*\natexlab[1]{#1}
\providecommand*\mciteSetBstSublistMode[1]{}
\providecommand*\mciteSetBstMaxWidthForm[2]{}
\providecommand*\mciteBstWouldAddEndPuncttrue
  {\def\EndOfBibitem{\unskip.}}
\providecommand*\mciteBstWouldAddEndPunctfalse
  {\let\EndOfBibitem\relax}
\providecommand*\mciteSetBstMidEndSepPunct[3]{}
\providecommand*\mciteSetBstSublistLabelBeginEnd[3]{}
\providecommand*\EndOfBibitem{}
\mciteSetBstSublistMode{f}
\mciteSetBstMaxWidthForm{subitem}{(\alph{mcitesubitemcount})}
\mciteSetBstSublistLabelBeginEnd
  {\mcitemaxwidthsubitemform\space}
  {\relax}
  {\relax}

\bibitem[Armitage \latin{et~al.}(2018)Armitage, Mele, and
  Vishwanath]{Armitage2018_WeylDirac_review}
Armitage,~N.~P.; Mele,~E.~J.; Vishwanath,~A. Weyl and {{Dirac}} semimetals in
  three-dimensional solids. \emph{Rev. Mod. Phys.} \textbf{2018}, \emph{90},
  015001\relax
\mciteBstWouldAddEndPuncttrue
\mciteSetBstMidEndSepPunct{\mcitedefaultmidpunct}
{\mcitedefaultendpunct}{\mcitedefaultseppunct}\relax
\EndOfBibitem
\bibitem[He \latin{et~al.}(2014)He, Hong, Dong, Pan, Zhang, Zhang, and
  Li]{He_Cd3As2_2014}
He,~L.~P.; Hong,~X.~C.; Dong,~J.~K.; Pan,~J.; Zhang,~Z.; Zhang,~J.; Li,~S.~Y.
  Quantum Transport Evidence for the Three-Dimensional {{Dirac}} Semimetal
  Phase in $\mathrm{Cd}_{3}\mathrm{As}_{2}$. \emph{Phys. Rev. Lett.}
  \textbf{2014}, \emph{113}, 246402\relax
\mciteBstWouldAddEndPuncttrue
\mciteSetBstMidEndSepPunct{\mcitedefaultmidpunct}
{\mcitedefaultendpunct}{\mcitedefaultseppunct}\relax
\EndOfBibitem
\bibitem[Liang \latin{et~al.}(2015)Liang, Gibson, Ali, Liu, Cava, and
  Ong]{Liang_Cd3As2_2015}
Liang,~T.; Gibson,~Q.; Ali,~M.~N.; Liu,~M.; Cava,~R.~J.; Ong,~N.~P. Ultrahigh
  mobility and giant magnetoresistance in the {{Dirac}} semimetal
  $\mathrm{Cd}_3\mathrm{As}_2$. \emph{Nat. Mater.} \textbf{2015}, \emph{14},
  280--284\relax
\mciteBstWouldAddEndPuncttrue
\mciteSetBstMidEndSepPunct{\mcitedefaultmidpunct}
{\mcitedefaultendpunct}{\mcitedefaultseppunct}\relax
\EndOfBibitem
\bibitem[Cao \latin{et~al.}(2015)Cao, Liang, Zhang, Liu, Huang, Jin, Chen,
  Wang, Wang, Zhao, Li, Dai, Zou, Xia, Li, and Xiu]{Cao_Cd3As2_2015}
Cao,~J. \latin{et~al.}  Landau level splitting in $\mathrm{Cd}_3\mathrm{As}_2$
  under high magnetic fields. \emph{Nat. Commun.} \textbf{2015}, \emph{6},
  7779\relax
\mciteBstWouldAddEndPuncttrue
\mciteSetBstMidEndSepPunct{\mcitedefaultmidpunct}
{\mcitedefaultendpunct}{\mcitedefaultseppunct}\relax
\EndOfBibitem
\bibitem[Feng \latin{et~al.}(2015)Feng, Pang, Wu, Wang, Weng, Li, Dai, Fang,
  Shi, and Lu]{FengCd3As2_2015}
Feng,~J.; Pang,~Y.; Wu,~D.; Wang,~Z.; Weng,~H.; Li,~J.; Dai,~X.; Fang,~Z.;
  Shi,~Y.; Lu,~L. Large linear magnetoresistance in {{Dirac}} semimetal
  $\mathrm{Cd}_{3}\mathrm{As}_{2}$ with Fermi surfaces close to the {{Dirac}}
  points. \emph{Phys. Rev. B} \textbf{2015}, \emph{92}, 081306\relax
\mciteBstWouldAddEndPuncttrue
\mciteSetBstMidEndSepPunct{\mcitedefaultmidpunct}
{\mcitedefaultendpunct}{\mcitedefaultseppunct}\relax
\EndOfBibitem
\bibitem[Narayanan \latin{et~al.}(2015)Narayanan, Watson, Blake, Bruyant,
  Drigo, Chen, Prabhakaran, Yan, Felser, Kong, Canfield, and
  Coldea]{Narayanan_Cd3As2_2015}
Narayanan,~A.; Watson,~M.~D.; Blake,~S.~F.; Bruyant,~N.; Drigo,~L.;
  Chen,~Y.~L.; Prabhakaran,~D.; Yan,~B.; Felser,~C.; Kong,~T.; Canfield,~P.~C.;
  Coldea,~A.~I. Linear Magnetoresistance Caused by Mobility Fluctuations in
  $n$-Doped $\mathrm{Cd}_{3}\mathrm{As}_{2}$. \emph{Phys. Rev. Lett.}
  \textbf{2015}, \emph{114}, 117201\relax
\mciteBstWouldAddEndPuncttrue
\mciteSetBstMidEndSepPunct{\mcitedefaultmidpunct}
{\mcitedefaultendpunct}{\mcitedefaultseppunct}\relax
\EndOfBibitem
\bibitem[Amarnath \latin{et~al.}(2019)Amarnath, Bhargavi, and
  Kubakaddi]{Amarnath_Cd3As2_2019}
Amarnath,~R.; Bhargavi,~K.~S.; Kubakaddi,~S.~S. Phonon limited mobility in 3{D}
  {Dirac} semimetal $\mathrm{Cd}_3\mathrm{As}_2$. \emph{IOP Conf. Ser.: Mater.
  Sci. Eng.} \textbf{2019}, \emph{561}, 012030\relax
\mciteBstWouldAddEndPuncttrue
\mciteSetBstMidEndSepPunct{\mcitedefaultmidpunct}
{\mcitedefaultendpunct}{\mcitedefaultseppunct}\relax
\EndOfBibitem
\bibitem[Xiong \latin{et~al.}(2015)Xiong, Kushwaha, Liang, Krizan,
  Hirschberger, Wang, Cava, and Ong]{Xiong_chiralNa3Bi_2015}
Xiong,~J.; Kushwaha,~S.~K.; Liang,~T.; Krizan,~J.~W.; Hirschberger,~M.;
  Wang,~W.; Cava,~R.~J.; Ong,~N.~P. Evidence for the chiral anomaly in the
  {Dirac} semimetal {Na$_3$Bi}. \emph{Science} \textbf{2015}, \emph{350},
  413--416\relax
\mciteBstWouldAddEndPuncttrue
\mciteSetBstMidEndSepPunct{\mcitedefaultmidpunct}
{\mcitedefaultendpunct}{\mcitedefaultseppunct}\relax
\EndOfBibitem
\bibitem[Xiong \latin{et~al.}(2016)Xiong, Kushwaha, Krizan, Liang, Cava, and
  Ong]{xiongAnomalous2016}
Xiong,~J.; Kushwaha,~S.; Krizan,~J.; Liang,~T.; Cava,~R.~J.; Ong,~N.~P.
  Anomalous Conductivity Tensor in the {Dirac} Semimetal
  $\mathrm{Na}_3\mathrm{Bi}$. \emph{Europhys. Lett.} \textbf{2016}, \emph{114},
  27002\relax
\mciteBstWouldAddEndPuncttrue
\mciteSetBstMidEndSepPunct{\mcitedefaultmidpunct}
{\mcitedefaultendpunct}{\mcitedefaultseppunct}\relax
\EndOfBibitem
\bibitem[Li \latin{et~al.}(2015)Li, Wang, Liu, Wang, Liao, and
  Yu]{Li2015_Cd3As2_chiralanomaly}
Li,~C.-Z.; Wang,~L.-X.; Liu,~H.; Wang,~J.; Liao,~Z.-M.; Yu,~D.-P. Giant
  negative magnetoresistance induced by the chiral anomaly in individual
  $\mathrm{Cd}_3\mathrm{As}_2$ nanowires. \emph{Nat. Commun.} \textbf{2015},
  \emph{6}, 10137\relax
\mciteBstWouldAddEndPuncttrue
\mciteSetBstMidEndSepPunct{\mcitedefaultmidpunct}
{\mcitedefaultendpunct}{\mcitedefaultseppunct}\relax
\EndOfBibitem
\bibitem[Wang \latin{et~al.}(2012)Wang, Sun, Chen, Franchini, Xu, Weng, Dai,
  and Fang]{ZhijunWang2012_Na3Bi}
Wang,~Z.; Sun,~Y.; Chen,~X.-Q.; Franchini,~C.; Xu,~G.; Weng,~H.; Dai,~X.;
  Fang,~Z. Dirac semimetal and topological phase transitions in {A}$_{3}${Bi}
  ({A}=$\text{Na}$, {K}, {Rb}). \emph{Phys. Rev. B} \textbf{2012}, \emph{85},
  195320\relax
\mciteBstWouldAddEndPuncttrue
\mciteSetBstMidEndSepPunct{\mcitedefaultmidpunct}
{\mcitedefaultendpunct}{\mcitedefaultseppunct}\relax
\EndOfBibitem
\bibitem[Liang \latin{et~al.}(2016)Liang, Chen, Wang, Shi, Feng, Yi, Xie, He,
  He, Peng, Liu, Liu, Hu, Zhao, Liu, Dong, Zhang, Nakatake, Iwasawa, Shimada,
  Arita, Namatame, Taniguchi, Xu, Chen, Weng, Dai, Fang, and
  Zhou]{LiangArpesNa3Bi_2016}
Liang,~A. \latin{et~al.}  Electronic structure, {Dirac} points and {Fermi} arc
  surface states in three-dimensional {Dirac} semimetal
  $\mathrm{Na}_3\mathrm{Bi}$ from angle-resolved photoemission spectroscopy.
  \emph{Chin. Phys. B} \textbf{2016}, \emph{25}, 077101\relax
\mciteBstWouldAddEndPuncttrue
\mciteSetBstMidEndSepPunct{\mcitedefaultmidpunct}
{\mcitedefaultendpunct}{\mcitedefaultseppunct}\relax
\EndOfBibitem
\bibitem[Xu \latin{et~al.}(2015)Xu, Liu, Kushwaha, Sankar, Krizan, Belopolski,
  Neupane, Bian, Alidoust, Chang, Jeng, Huang, Tsai, Lin, Shibayev, Chou, Cava,
  and Hasan]{Xu_Na3Bi_Fermiarcs2015}
Xu,~S.-Y. \latin{et~al.}  Observation of {Fermi} arc surface states in a
  topological metal. \emph{Science} \textbf{2015}, \emph{347}, 294--298\relax
\mciteBstWouldAddEndPuncttrue
\mciteSetBstMidEndSepPunct{\mcitedefaultmidpunct}
{\mcitedefaultendpunct}{\mcitedefaultseppunct}\relax
\EndOfBibitem
\bibitem[Moll \latin{et~al.}(2016)Moll, Nair, Helm, Potter, Kimchi, Vishwanath,
  and Analytis]{Moll2016_Cd3As2_fermiarcs}
Moll,~P. J.~W.; Nair,~N.~L.; Helm,~T.; Potter,~A.~C.; Kimchi,~I.;
  Vishwanath,~A.; Analytis,~J.~G. Transport evidence for {Fermi}-arc-mediated
  chirality transfer in the {Dirac} semimetal $\mathrm{Cd}_3\mathrm{As}_2$.
  \emph{Nature} \textbf{2016}, \emph{535}, 266--270\relax
\mciteBstWouldAddEndPuncttrue
\mciteSetBstMidEndSepPunct{\mcitedefaultmidpunct}
{\mcitedefaultendpunct}{\mcitedefaultseppunct}\relax
\EndOfBibitem
\bibitem[Collins \latin{et~al.}(2018)Collins, Tadich, Wu, Gomes, Rodrigues,
  Liu, Hellerstedt, Ryu, Tang, Mo, Adam, Yang, Fuhrer, and
  Edmonds]{Collins2018_Na3bitransistor}
Collins,~J.~L.; Tadich,~A.; Wu,~W.; Gomes,~L.~C.; Rodrigues,~J. N.~B.; Liu,~C.;
  Hellerstedt,~J.; Ryu,~H.; Tang,~S.; Mo,~S.-K.; Adam,~S.; Yang,~S.~A.;
  Fuhrer,~M.~S.; Edmonds,~M.~T. Electric-field-tuned topological phase
  transition in ultrathin {{Na$_3$Bi}}. \emph{Nature} \textbf{2018},
  \emph{564}, 390--394\relax
\mciteBstWouldAddEndPuncttrue
\mciteSetBstMidEndSepPunct{\mcitedefaultmidpunct}
{\mcitedefaultendpunct}{\mcitedefaultseppunct}\relax
\EndOfBibitem
\bibitem[Xia \latin{et~al.}(2019)Xia, Li, Cai, Qin, Zou, Peng, Duan, Xu, Zhang,
  and Fu]{Xia_Na3BiSTM_2019}
Xia,~H.; Li,~Y.; Cai,~M.; Qin,~L.; Zou,~N.; Peng,~L.; Duan,~W.; Xu,~Y.;
  Zhang,~W.; Fu,~Y.-S. Dimensional Crossover and Topological Phase Transition
  in {Dirac} Semimetal {{Na$_3$Bi}} Films. \emph{ACS Nano} \textbf{2019},
  \emph{13}, 9647--9654\relax
\mciteBstWouldAddEndPuncttrue
\mciteSetBstMidEndSepPunct{\mcitedefaultmidpunct}
{\mcitedefaultendpunct}{\mcitedefaultseppunct}\relax
\EndOfBibitem
\bibitem[Geim and Novoselov(2007)Geim, and Novoselov]{geim2007rise}
Geim,~A.~K.; Novoselov,~K.~S. The rise of graphene. \emph{Nat. Mater.}
  \textbf{2007}, \emph{6}, 183\relax
\mciteBstWouldAddEndPuncttrue
\mciteSetBstMidEndSepPunct{\mcitedefaultmidpunct}
{\mcitedefaultendpunct}{\mcitedefaultseppunct}\relax
\EndOfBibitem
\bibitem[Hills \latin{et~al.}(2017)Hills, Kusmartseva, and
  Kusmartsev]{Dirac_appl_2017}
Hills,~R. D.~Y.; Kusmartseva,~A.; Kusmartsev,~F.~V. Current-voltage
  characteristics of Weyl semimetal semiconducting devices, Veselago lenses,
  and hyperbolic Dirac phase. \emph{Phys. Rev. B} \textbf{2017}, \emph{95},
  214103\relax
\mciteBstWouldAddEndPuncttrue
\mciteSetBstMidEndSepPunct{\mcitedefaultmidpunct}
{\mcitedefaultendpunct}{\mcitedefaultseppunct}\relax
\EndOfBibitem
\bibitem[Yanez \latin{et~al.}(2021)Yanez, Ou, Xiao, Koo, Held, Ghosh, Rable,
  Pillsbury, Delgado, Yang, Chamorro, Grutter, Quarterman, Richardella,
  Sengupta, McQueen, Borchers, Mkhoyan, Yan, and Samarth]{spin_charge_2021}
Yanez,~W. \latin{et~al.}  Spin and Charge Interconversion in Dirac-Semimetal
  Thin Films. \emph{Phys. Rev. Appl.} \textbf{2021}, \emph{16}, 054031\relax
\mciteBstWouldAddEndPuncttrue
\mciteSetBstMidEndSepPunct{\mcitedefaultmidpunct}
{\mcitedefaultendpunct}{\mcitedefaultseppunct}\relax
\EndOfBibitem
\bibitem[Sie \latin{et~al.}(2019)Sie, Nyby, Pemmaraju, Park, Shen, Yang,
  Hoffmann, Ofori-Okai, Li, Reid, Weathersby, Mannebach, Finney, Rhodes,
  Chenet, Antony, Balicas, Hone, Devereaux, Heinz, Wang, and
  Lindenberg]{Sie2019}
Sie,~E.~J. \latin{et~al.}  An ultrafast symmetry switch in a Weyl semimetal.
  \emph{Nature} \textbf{2019}, \emph{565}, 61--66\relax
\mciteBstWouldAddEndPuncttrue
\mciteSetBstMidEndSepPunct{\mcitedefaultmidpunct}
{\mcitedefaultendpunct}{\mcitedefaultseppunct}\relax
\EndOfBibitem
\bibitem[Zhang \latin{et~al.}(2019)Zhang, Wang, Li, Shi, Wu, Lin, Zhang, Liu,
  Liu, Wang, Dong, and Wang]{Zhang_MoTe2_2019}
Zhang,~M.~Y.; Wang,~Z.~X.; Li,~Y.~N.; Shi,~L.~Y.; Wu,~D.; Lin,~T.;
  Zhang,~S.~J.; Liu,~Y.~Q.; Liu,~Q.~M.; Wang,~J.; Dong,~T.; Wang,~N.~L.
  Light-Induced Subpicosecond Lattice Symmetry Switch in ${\mathrm{MoTe}}_{2}$.
  \emph{Phys. Rev. X} \textbf{2019}, \emph{9}, 021036\relax
\mciteBstWouldAddEndPuncttrue
\mciteSetBstMidEndSepPunct{\mcitedefaultmidpunct}
{\mcitedefaultendpunct}{\mcitedefaultseppunct}\relax
\EndOfBibitem
\bibitem[Vaswani \latin{et~al.}(2020)Vaswani, Wang, Mudiyanselage, Li, Lozano,
  Gu, Cheng, Song, Luo, Kim, Huang, Liu, Mootz, Perakis, Yao, Ho, and
  Wang]{Vaswani_2020}
Vaswani,~C. \latin{et~al.}  Light-Driven Raman Coherence as a Nonthermal Route
  to Ultrafast Topology Switching in a Dirac Semimetal. \emph{Phys. Rev. X}
  \textbf{2020}, \emph{10}, 021013\relax
\mciteBstWouldAddEndPuncttrue
\mciteSetBstMidEndSepPunct{\mcitedefaultmidpunct}
{\mcitedefaultendpunct}{\mcitedefaultseppunct}\relax
\EndOfBibitem
\bibitem[Xiao \latin{et~al.}(2020)Xiao, Wang, Wang, Pemmaraju, Wang, Muscher,
  Sie, Nyby, Devereaux, Qian, Zhang, and Lindenberg]{Xiao2020}
Xiao,~J.; Wang,~Y.; Wang,~H.; Pemmaraju,~C.~D.; Wang,~S.; Muscher,~P.;
  Sie,~E.~J.; Nyby,~C.~M.; Devereaux,~T.~P.; Qian,~X.; Zhang,~X.;
  Lindenberg,~A.~M. Berry curvature memory through electrically driven stacking
  transitions. \emph{Nat. Phys.} \textbf{2020}, \emph{16}, 1028\relax
\mciteBstWouldAddEndPuncttrue
\mciteSetBstMidEndSepPunct{\mcitedefaultmidpunct}
{\mcitedefaultendpunct}{\mcitedefaultseppunct}\relax
\EndOfBibitem
\bibitem[Disa \latin{et~al.}(2021)Disa, Nova, and Cavalleri]{Disa2021}
Disa,~A.~S.; Nova,~T.~F.; Cavalleri,~A. Engineering crystal structures with
  light. \emph{Nat. Phys.} \textbf{2021}, \emph{17}, 1087--1092\relax
\mciteBstWouldAddEndPuncttrue
\mciteSetBstMidEndSepPunct{\mcitedefaultmidpunct}
{\mcitedefaultendpunct}{\mcitedefaultseppunct}\relax
\EndOfBibitem
\bibitem[Sklyadneva \latin{et~al.}(2021)Sklyadneva, Heid, Echenique, and
  Chulkov]{Yu_zrcosn_2021}
Sklyadneva,~I.; Heid,~R.; Echenique,~P.~M.; Chulkov,~E.~V. Electron-phonon
  coupling in the magnetic Weyl semimetal
  $\mathrm{Zr}{\mathrm{Co}}_{2}\mathrm{Sn}$. \emph{Phys. Rev. B} \textbf{2021},
  \emph{103}, 024303\relax
\mciteBstWouldAddEndPuncttrue
\mciteSetBstMidEndSepPunct{\mcitedefaultmidpunct}
{\mcitedefaultendpunct}{\mcitedefaultseppunct}\relax
\EndOfBibitem
\bibitem[Liu \latin{et~al.}(2021)Liu, Li, Zhang, Tu, Zhang, Zhang, Wang, and
  Singh]{Liu_biphenylene_2021}
Liu,~P.-F.; Li,~J.; Zhang,~C.; Tu,~X.-H.; Zhang,~J.; Zhang,~P.; Wang,~B.-T.;
  Singh,~D.~J. Type-II Dirac cones and electron-phonon interaction in monolayer
  biphenylene from first-principles calculations. \emph{Phys. Rev. B}
  \textbf{2021}, \emph{104}, 235422\relax
\mciteBstWouldAddEndPuncttrue
\mciteSetBstMidEndSepPunct{\mcitedefaultmidpunct}
{\mcitedefaultendpunct}{\mcitedefaultseppunct}\relax
\EndOfBibitem
\bibitem[Bernardi(2016)]{Bernardi_2016}
Bernardi,~M. First-principles dynamics of electrons and phonons. \emph{Eur.
  Phys. J. B} \textbf{2016}, \emph{89}, 239\relax
\mciteBstWouldAddEndPuncttrue
\mciteSetBstMidEndSepPunct{\mcitedefaultmidpunct}
{\mcitedefaultendpunct}{\mcitedefaultseppunct}\relax
\EndOfBibitem
\bibitem[Agapito and Bernardi(2018)Agapito, and Bernardi]{Agapito_2018}
Agapito,~L.~A.; Bernardi,~M. Ab initio electron-phonon interactions using
  atomic orbital wave functions. \emph{Phys. Rev. B} \textbf{2018}, \emph{97},
  235146\relax
\mciteBstWouldAddEndPuncttrue
\mciteSetBstMidEndSepPunct{\mcitedefaultmidpunct}
{\mcitedefaultendpunct}{\mcitedefaultseppunct}\relax
\EndOfBibitem
\bibitem[Bernardi \latin{et~al.}(2014)Bernardi, Vigil-Fowler, Lischner, Neaton,
  and Louie]{Bernardi_Si_2014}
Bernardi,~M.; Vigil-Fowler,~D.; Lischner,~J.; Neaton,~J.~B.; Louie,~S.~G. Ab
  Initio Study of Hot Carriers in the First Picosecond after Sunlight
  Absorption in Silicon. \emph{Phys. Rev. Lett.} \textbf{2014}, \emph{112},
  257402\relax
\mciteBstWouldAddEndPuncttrue
\mciteSetBstMidEndSepPunct{\mcitedefaultmidpunct}
{\mcitedefaultendpunct}{\mcitedefaultseppunct}\relax
\EndOfBibitem
\bibitem[Zhou and Bernardi(2016)Zhou, and Bernardi]{Zhou2016}
Zhou,~J.-J.; Bernardi,~M. Ab initio electron mobility and polar phonon
  scattering in {G}a{A}s. \emph{Phys. Rev. B} \textbf{2016}, \emph{94},
  201201(R)\relax
\mciteBstWouldAddEndPuncttrue
\mciteSetBstMidEndSepPunct{\mcitedefaultmidpunct}
{\mcitedefaultendpunct}{\mcitedefaultseppunct}\relax
\EndOfBibitem
\bibitem[Zhou \latin{et~al.}(2018)Zhou, Hellman, and Bernardi]{Zhou_STO_2018}
Zhou,~J.-J.; Hellman,~O.; Bernardi,~M. Electron-Phonon Scattering in the
  Presence of Soft Modes and Electron Mobility in ${\mathrm{SrTiO}}_{3}$
  Perovskite from First Principles. \emph{Phys. Rev. Lett.} \textbf{2018},
  \emph{121}, 226603\relax
\mciteBstWouldAddEndPuncttrue
\mciteSetBstMidEndSepPunct{\mcitedefaultmidpunct}
{\mcitedefaultendpunct}{\mcitedefaultseppunct}\relax
\EndOfBibitem
\bibitem[Zhou and Bernardi(2019)Zhou, and Bernardi]{Zhou_STO_2019}
Zhou,~J.-J.; Bernardi,~M. Predicting charge transport in the presence of
  polarons: The beyond-quasiparticle regime in ${\mathrm{SrTiO}}_{3}$.
  \emph{Phys. Rev. Res.} \textbf{2019}, \emph{1}, 033138\relax
\mciteBstWouldAddEndPuncttrue
\mciteSetBstMidEndSepPunct{\mcitedefaultmidpunct}
{\mcitedefaultendpunct}{\mcitedefaultseppunct}\relax
\EndOfBibitem
\bibitem[Park \latin{et~al.}(2007)Park, Giustino, Cohen, and
  Louie]{Park_graphene_2007}
Park,~C.-H.; Giustino,~F.; Cohen,~M.~L.; Louie,~S.~G. Velocity Renormalization
  and Carrier Lifetime in Graphene from the Electron-Phonon Interaction.
  \emph{Phys. Rev. Lett.} \textbf{2007}, \emph{99}, 086804\relax
\mciteBstWouldAddEndPuncttrue
\mciteSetBstMidEndSepPunct{\mcitedefaultmidpunct}
{\mcitedefaultendpunct}{\mcitedefaultseppunct}\relax
\EndOfBibitem
\bibitem[Floris \latin{et~al.}(2007)Floris, Sanna, Massidda, and
  Gross]{Floris_Pb_2007}
Floris,~A.; Sanna,~A.; Massidda,~S.; Gross,~E. K.~U. Two-band superconductivity
  in Pb from ab initio calculations. \emph{Phys. Rev. B} \textbf{2007},
  \emph{75}, 054508\relax
\mciteBstWouldAddEndPuncttrue
\mciteSetBstMidEndSepPunct{\mcitedefaultmidpunct}
{\mcitedefaultendpunct}{\mcitedefaultseppunct}\relax
\EndOfBibitem
\bibitem[Li(2015)]{Li2015}
Li,~W. Electrical transport limited by electron-phonon coupling from
  {Boltzmann} transport equation: An ab initio study of {S}i, {A}l, and
  {M}o{S}$_{\mathrm{2}}$. \emph{Phys. Rev. B} \textbf{2015}, \emph{92},
  075405\relax
\mciteBstWouldAddEndPuncttrue
\mciteSetBstMidEndSepPunct{\mcitedefaultmidpunct}
{\mcitedefaultendpunct}{\mcitedefaultseppunct}\relax
\EndOfBibitem
\bibitem[Liu \latin{et~al.}(2017)Liu, Zhou, Liao, Singh, and Chen]{Chen2017}
Liu,~T.-H.; Zhou,~J.; Liao,~B.; Singh,~D.~J.; Chen,~G. First-principles
  mode-by-mode analysis for electron-phonon scattering channels and mean free
  path spectra in {G}a{A}s. \emph{Phys. Rev. B} \textbf{2017}, \emph{95},
  075206\relax
\mciteBstWouldAddEndPuncttrue
\mciteSetBstMidEndSepPunct{\mcitedefaultmidpunct}
{\mcitedefaultendpunct}{\mcitedefaultseppunct}\relax
\EndOfBibitem
\bibitem[Ma \latin{et~al.}(2018)Ma, Nissimagoudar, and Li]{Li2018}
Ma,~J.; Nissimagoudar,~A.~S.; Li,~W. First-principles study of electron and
  hole mobilities of {S}i and {G}a{A}s. \emph{Phys. Rev. B} \textbf{2018},
  \emph{97}, 045201\relax
\mciteBstWouldAddEndPuncttrue
\mciteSetBstMidEndSepPunct{\mcitedefaultmidpunct}
{\mcitedefaultendpunct}{\mcitedefaultseppunct}\relax
\EndOfBibitem
\bibitem[Ponc\'e \latin{et~al.}(2018)Ponc\'e, Margine, and
  Giustino]{Giustino2018}
Ponc\'e,~S.; Margine,~E.~R.; Giustino,~F. Towards predictive many-body
  calculations of phonon-limited carrier mobilities in semiconductors.
  \emph{Phys. Rev. B} \textbf{2018}, \emph{97}, 121201\relax
\mciteBstWouldAddEndPuncttrue
\mciteSetBstMidEndSepPunct{\mcitedefaultmidpunct}
{\mcitedefaultendpunct}{\mcitedefaultseppunct}\relax
\EndOfBibitem
\bibitem[Liu \latin{et~al.}(2014)Liu, Zhou, Zhang, Wang, Weng, Prabhakaran, Mo,
  Shen, Fang, Dai, Hussain, and Chen]{liuDiscovery2014}
Liu,~Z.~K.; Zhou,~B.; Zhang,~Y.; Wang,~Z.~J.; Weng,~H.~M.; Prabhakaran,~D.;
  Mo,~S.-K.; Shen,~Z.~X.; Fang,~Z.; Dai,~X.; Hussain,~Z.; Chen,~Y.~L. Discovery
  of a {{Three}}-{{Dimensional Topological Dirac Semimetal}}, {{Na$_3$Bi}}.
  \emph{Science} \textbf{2014}, \emph{343}, 864--867\relax
\mciteBstWouldAddEndPuncttrue
\mciteSetBstMidEndSepPunct{\mcitedefaultmidpunct}
{\mcitedefaultendpunct}{\mcitedefaultseppunct}\relax
\EndOfBibitem
\bibitem[Kushwaha \latin{et~al.}(2015)Kushwaha, Krizan, Feldman, Gyenis,
  Randeria, Xiong, Xu, Alidoust, Belopolski, Liang, Zahid~Hasan, Ong, Yazdani,
  and Cava]{kushwahaBulk2015}
Kushwaha,~S.~K.; Krizan,~J.~W.; Feldman,~B.~E.; Gyenis,~A.; Randeria,~M.~T.;
  Xiong,~J.; Xu,~S.-Y.; Alidoust,~N.; Belopolski,~I.; Liang,~T.;
  Zahid~Hasan,~M.; Ong,~N.~P.; Yazdani,~A.; Cava,~R.~J. Bulk Crystal Growth and
  Electronic Characterization of the {{3D Dirac}} Semimetal {{Na$_3$Bi}}.
  \emph{APL Mater.} \textbf{2015}, \emph{3}, 041504\relax
\mciteBstWouldAddEndPuncttrue
\mciteSetBstMidEndSepPunct{\mcitedefaultmidpunct}
{\mcitedefaultendpunct}{\mcitedefaultseppunct}\relax
\EndOfBibitem
\bibitem[Edmonds \latin{et~al.}(2017)Edmonds, Collins, Hellerstedt, Yudhistira,
  Gomes, Rodrigues, Adam, and Fuhrer]{Edmonds_Na3BiSTM_2017}
Edmonds,~M.~T.; Collins,~J.~L.; Hellerstedt,~J.; Yudhistira,~I.; Gomes,~L.~C.;
  Rodrigues,~J. N.~B.; Adam,~S.; Fuhrer,~M.~S. Spatial charge inhomogeneity and
  defect states in topological {Dirac} semimetal thin films of {{Na$_3$Bi}}.
  \emph{Sci. Adv.} \textbf{2017}, \emph{3}, eaao6661\relax
\mciteBstWouldAddEndPuncttrue
\mciteSetBstMidEndSepPunct{\mcitedefaultmidpunct}
{\mcitedefaultendpunct}{\mcitedefaultseppunct}\relax
\EndOfBibitem
\bibitem[Di~Bernardo \latin{et~al.}(2020)Di~Bernardo, Collins, Wu, Zhou, Yang,
  Ju, Edmonds, and Fuhrer]{dibernardoImportance2020}
Di~Bernardo,~I.; Collins,~J.; Wu,~W.; Zhou,~J.; Yang,~S.~A.; Ju,~S.;
  Edmonds,~M.~T.; Fuhrer,~M.~S. Importance of Interactions for the Band
  Structure of the Topological {{Dirac}} Semimetal {{Na$_3$Bi}}. \emph{Phys.
  Rev. B} \textbf{2020}, \emph{102}, 045124\relax
\mciteBstWouldAddEndPuncttrue
\mciteSetBstMidEndSepPunct{\mcitedefaultmidpunct}
{\mcitedefaultendpunct}{\mcitedefaultseppunct}\relax
\EndOfBibitem
\bibitem[Yuan \latin{et~al.}(2019)Yuan, Xu, Zhao, Song, Zhang, Xiao, Ding, and
  Peeters]{YuanQuantum2019}
Yuan,~H.~F.; Xu,~W.; Zhao,~X.~N.; Song,~D.; Zhang,~G.~R.; Xiao,~Y.~M.;
  Ding,~L.; Peeters,~F.~M. Quantum and transport mobilities of a
  $\mathrm{Na}_{3}\mathrm{Bi}$-based three-dimensional {Dirac} system.
  \emph{Phys. Rev. B} \textbf{2019}, \emph{99}, 235303\relax
\mciteBstWouldAddEndPuncttrue
\mciteSetBstMidEndSepPunct{\mcitedefaultmidpunct}
{\mcitedefaultendpunct}{\mcitedefaultseppunct}\relax
\EndOfBibitem
\bibitem[H{\"u}bener \latin{et~al.}(2017)H{\"u}bener, Sentef, De~Giovannini,
  Kemper, and Rubio]{Hubener2017}
H{\"u}bener,~H.; Sentef,~M.~A.; De~Giovannini,~U.; Kemper,~A.~F.; Rubio,~A.
  Creating stable Floquet--Weyl semimetals by laser-driving of 3D Dirac
  materials. \emph{Nat. Commun.} \textbf{2017}, \emph{8}, 13940\relax
\mciteBstWouldAddEndPuncttrue
\mciteSetBstMidEndSepPunct{\mcitedefaultmidpunct}
{\mcitedefaultendpunct}{\mcitedefaultseppunct}\relax
\EndOfBibitem
\bibitem[Tancogne-Dejean \latin{et~al.}(2022)Tancogne-Dejean, Eich, and
  Rubio]{Tancogne-Dejean2022}
Tancogne-Dejean,~N.; Eich,~F.~G.; Rubio,~A. Effect of spin-orbit coupling on
  the high harmonics from the topological Dirac semimetal Na$_3$Bi. \emph{npj
  Comput. Mater.} \textbf{2022}, \emph{8}, 145\relax
\mciteBstWouldAddEndPuncttrue
\mciteSetBstMidEndSepPunct{\mcitedefaultmidpunct}
{\mcitedefaultendpunct}{\mcitedefaultseppunct}\relax
\EndOfBibitem
\bibitem[Martin(2004)]{Martin-book}
Martin,~R.~M. \emph{Electronic Structure: Basic Theory and Practical Methods};
  Cambridge University Press, 2004\relax
\mciteBstWouldAddEndPuncttrue
\mciteSetBstMidEndSepPunct{\mcitedefaultmidpunct}
{\mcitedefaultendpunct}{\mcitedefaultseppunct}\relax
\EndOfBibitem
\bibitem[Baroni \latin{et~al.}(2001)Baroni, de~Gironcoli, Dal~Corso, and
  Giannozzi]{Baroni_DFPT_2001}
Baroni,~S.; de~Gironcoli,~S.; Dal~Corso,~A.; Giannozzi,~P. Phonons and related
  crystal properties from density-functional perturbation theory. \emph{Rev.
  Mod. Phys.} \textbf{2001}, \emph{73}, 515--562\relax
\mciteBstWouldAddEndPuncttrue
\mciteSetBstMidEndSepPunct{\mcitedefaultmidpunct}
{\mcitedefaultendpunct}{\mcitedefaultseppunct}\relax
\EndOfBibitem
\bibitem[Hybertsen and Louie(1986)Hybertsen, and Louie]{Hybertsen_GW_1986}
Hybertsen,~M.~S.; Louie,~S.~G. Electron correlation in semiconductors and
  insulators: Band gaps and quasiparticle energies. \emph{Phys. Rev. B}
  \textbf{1986}, \emph{34}, 5390--5413\relax
\mciteBstWouldAddEndPuncttrue
\mciteSetBstMidEndSepPunct{\mcitedefaultmidpunct}
{\mcitedefaultendpunct}{\mcitedefaultseppunct}\relax
\EndOfBibitem
\bibitem[Schweicher \latin{et~al.}(2019)Schweicher, D'Avino, Ruggiero, Harkin,
  Broch, Venkateshvaran, Liu, Richard, Ruzié, Armstrong, Kennedy, Shankland,
  Takimiya, Geerts, Zeitler, Fratini, and Sirringhaus]{killer-ph}
Schweicher,~G. \latin{et~al.}  Chasing the “Killer” Phonon Mode for the
  Rational Design of Low-Disorder, High-Mobility Molecular Semiconductors.
  \emph{Adv. Mater.} \textbf{2019}, \emph{31}, 1902407\relax
\mciteBstWouldAddEndPuncttrue
\mciteSetBstMidEndSepPunct{\mcitedefaultmidpunct}
{\mcitedefaultendpunct}{\mcitedefaultseppunct}\relax
\EndOfBibitem
\bibitem[Jenkins \latin{et~al.}(2016)Jenkins, Lane, Barbiellini, Sushkov,
  Carey, Liu, Krizan, Kushwaha, Gibson, Chang, Jeng, Lin, Cava, Bansil, and
  Drew]{Jenkins_Na3Bi_2016}
Jenkins,~G.~S.; Lane,~C.; Barbiellini,~B.; Sushkov,~A.~B.; Carey,~R.~L.;
  Liu,~F.; Krizan,~J.~W.; Kushwaha,~S.~K.; Gibson,~Q.; Chang,~T.-R.;
  Jeng,~H.-T.; Lin,~H.; Cava,~R.~J.; Bansil,~A.; Drew,~H.~D. Three-dimensional
  {Dirac} cone carrier dynamics in $\mathrm{Na}_{3}\mathrm{Bi}$ and
  $\mathrm{Cd}_{3}\mathrm{As}_{2}$. \emph{Phys. Rev. B} \textbf{2016},
  \emph{94}, 085121\relax
\mciteBstWouldAddEndPuncttrue
\mciteSetBstMidEndSepPunct{\mcitedefaultmidpunct}
{\mcitedefaultendpunct}{\mcitedefaultseppunct}\relax
\EndOfBibitem
\bibitem[Shao \latin{et~al.}(2017)Shao, Ruan, Wu, Chen, Guo, Zhang, Sun, Sheng,
  and Xing]{Shao_Na3Bibs_2017}
Shao,~D.; Ruan,~J.; Wu,~J.; Chen,~T.; Guo,~Z.; Zhang,~H.; Sun,~J.; Sheng,~L.;
  Xing,~D. Strain-induced quantum topological phase transitions in
  $\mathrm{Na}_{3}\mathrm{Bi}$. \emph{Phys. Rev. B} \textbf{2017}, \emph{96},
  075112\relax
\mciteBstWouldAddEndPuncttrue
\mciteSetBstMidEndSepPunct{\mcitedefaultmidpunct}
{\mcitedefaultendpunct}{\mcitedefaultseppunct}\relax
\EndOfBibitem
\bibitem[Chiu \latin{et~al.}(2020)Chiu, Singh, Mardanya, Nokelainen, Agarwal,
  Lin, Lane, Pussi, Barbiellini, and Bansil]{Chiu_Na3Bi_2020}
Chiu,~W.-C.; Singh,~B.; Mardanya,~S.; Nokelainen,~J.; Agarwal,~A.; Lin,~H.;
  Lane,~C.; Pussi,~K.; Barbiellini,~B.; Bansil,~A. Topological {Dirac}
  Semimetal Phase in {Bismuth} Based Anode Materials for {Sodium}-Ion
  Batteries. \emph{Condens. Matter} \textbf{2020}, \emph{5}, 39\relax
\mciteBstWouldAddEndPuncttrue
\mciteSetBstMidEndSepPunct{\mcitedefaultmidpunct}
{\mcitedefaultendpunct}{\mcitedefaultseppunct}\relax
\EndOfBibitem
\bibitem[Mounet(2005)]{Mounet_asr_2005}
Mounet,~N. Structural, vibrational and thermodynamic properties of carbon
  allotropes from first-principles: diamond, graphite, and nanotubes.
  \emph{Masters thesis} \textbf{2005}, \relax
\mciteBstWouldAddEndPunctfalse
\mciteSetBstMidEndSepPunct{\mcitedefaultmidpunct}
{}{\mcitedefaultseppunct}\relax
\EndOfBibitem
\bibitem[Giannozzi \latin{et~al.}(2009)Giannozzi, Baroni, Bonini, Calandra,
  Car, Cavazzoni, Ceresoli, Chiarotti, Cococcioni, Dabo, Corso, de~Gironcoli,
  Fabris, Fratesi, Gebauer, Gerstmann, Gougoussis, Kokalj, Lazzeri,
  Martin-Samos, Marzari, Mauri, Mazzarello, Paolini, Pasquarello, Paulatto,
  Sbraccia, Scandolo, Sclauzero, Seitsonen, Smogunov, Umari, and
  Wentzcovitch]{Giannozzi_QE_2009}
Giannozzi,~P. \latin{et~al.}  {QUANTUM} {ESPRESSO}: a modular and open-source
  software project for quantum simulations of materials. \emph{J. Phys.
  Condens. Matter} \textbf{2009}, \emph{21}, 395502\relax
\mciteBstWouldAddEndPuncttrue
\mciteSetBstMidEndSepPunct{\mcitedefaultmidpunct}
{\mcitedefaultendpunct}{\mcitedefaultseppunct}\relax
\EndOfBibitem
\bibitem[Cheng \latin{et~al.}(2014)Cheng, Li, Sun, Chen, Li, and
  Li]{chengGroundstate2014}
Cheng,~X.; Li,~R.; Sun,~Y.; Chen,~X.-Q.; Li,~D.; Li,~Y. Ground-State Phase in
  the Three-Dimensional Topological {{Dirac}} Semimetal {{Na$_3$Bi}}.
  \emph{Phys. Rev. B} \textbf{2014}, \emph{89}, 245201\relax
\mciteBstWouldAddEndPuncttrue
\mciteSetBstMidEndSepPunct{\mcitedefaultmidpunct}
{\mcitedefaultendpunct}{\mcitedefaultseppunct}\relax
\EndOfBibitem
\bibitem[Zhou \latin{et~al.}(2021)Zhou, Park, Lu, Maliyov, Tong, and
  Bernardi]{zhouPerturbo2020}
Zhou,~J.-J.; Park,~J.; Lu,~I.-T.; Maliyov,~I.; Tong,~X.; Bernardi,~M. Perturbo:
  A software package for ab initio electron–phonon interactions, charge
  transport and ultrafast dynamics. \emph{Comput. Phys. Commun.} \textbf{2021},
  \emph{264}, 107970\relax
\mciteBstWouldAddEndPuncttrue
\mciteSetBstMidEndSepPunct{\mcitedefaultmidpunct}
{\mcitedefaultendpunct}{\mcitedefaultseppunct}\relax
\EndOfBibitem
\bibitem[Pizzi \latin{et~al.}(2021)Pizzi, Milana, Ferrari, Marzari, and
  Gibertini]{Pizzi_shearbreathing_2021}
Pizzi,~G.; Milana,~S.; Ferrari,~A.~C.; Marzari,~N.; Gibertini,~M. Shear and
  Breathing Modes of Layered Materials. \emph{ACS Nano} \textbf{2021},
  \emph{15}, 12509--12534\relax
\mciteBstWouldAddEndPuncttrue
\mciteSetBstMidEndSepPunct{\mcitedefaultmidpunct}
{\mcitedefaultendpunct}{\mcitedefaultseppunct}\relax
\EndOfBibitem
\bibitem[Neupane \latin{et~al.}(2014)Neupane, Xu, Sankar, Alidoust, Bian, Liu,
  Belopolski, Chang, Jeng, Lin, Bansil, Chou, and Hasan]{Neupane_Cd3As2_2014}
Neupane,~M.; Xu,~S.-Y.; Sankar,~R.; Alidoust,~N.; Bian,~G.; Liu,~C.;
  Belopolski,~I.; Chang,~T.-R.; Jeng,~H.-T.; Lin,~H.; Bansil,~A.; Chou,~F.;
  Hasan,~M.~Z. Observation of a three-dimensional topological Dirac semimetal
  phase in high-mobility Cd3As2. \emph{Nature Communications} \textbf{2014},
  \emph{5}, 3786\relax
\mciteBstWouldAddEndPuncttrue
\mciteSetBstMidEndSepPunct{\mcitedefaultmidpunct}
{\mcitedefaultendpunct}{\mcitedefaultseppunct}\relax
\EndOfBibitem
\bibitem[Park \latin{et~al.}(2014)Park, Bonini, Sohier, Samsonidze, Kozinsky,
  Calandra, Mauri, and Marzari]{Park_graphene_2014}
Park,~C.-H.; Bonini,~N.; Sohier,~T.; Samsonidze,~G.; Kozinsky,~B.;
  Calandra,~M.; Mauri,~F.; Marzari,~N. Electron--Phonon Interactions and the
  Intrinsic Electrical Resistivity of Graphene. \emph{Nano Lett.}
  \textbf{2014}, \emph{14}, 1113--1119\relax
\mciteBstWouldAddEndPuncttrue
\mciteSetBstMidEndSepPunct{\mcitedefaultmidpunct}
{\mcitedefaultendpunct}{\mcitedefaultseppunct}\relax
\EndOfBibitem
\bibitem[Tong and Bernardi(2021)Tong, and Bernardi]{Xiao_graphene_2021}
Tong,~X.; Bernardi,~M. Toward precise simulations of the coupled ultrafast
  dynamics of electrons and atomic vibrations in materials. \emph{Phys. Rev.
  Res.} \textbf{2021}, \emph{3}, 023072\relax
\mciteBstWouldAddEndPuncttrue
\mciteSetBstMidEndSepPunct{\mcitedefaultmidpunct}
{\mcitedefaultendpunct}{\mcitedefaultseppunct}\relax
\EndOfBibitem
\bibitem[Desai \latin{et~al.}(2021)Desai, Zviazhynski, Zhou, and
  Bernardi]{desaiMagnetotransport2021}
Desai,~D.~C.; Zviazhynski,~B.; Zhou,~J.-J.; Bernardi,~M. Magnetotransport in
  semiconductors and two-dimensional materials from first principles.
  \emph{Phys. Rev. B} \textbf{2021}, \emph{103}, L161103\relax
\mciteBstWouldAddEndPuncttrue
\mciteSetBstMidEndSepPunct{\mcitedefaultmidpunct}
{\mcitedefaultendpunct}{\mcitedefaultseppunct}\relax
\EndOfBibitem
\bibitem[Dong \latin{et~al.}(2019)Dong, Chen, Wang, Lv, and
  Wang]{Dong_Na3BiPhonons_2019}
Dong,~X.-X.; Chen,~J.-X.; Wang,~Y.; Lv,~Z.-L.; Wang,~H.-Y. Electronic, elastic
  and lattice dynamic properties of the topological {Dirac} semimetal
  {{Na$_3$Bi}}. \emph{Mater. Res. Express} \textbf{2019}, \emph{6},
  076308\relax
\mciteBstWouldAddEndPuncttrue
\mciteSetBstMidEndSepPunct{\mcitedefaultmidpunct}
{\mcitedefaultendpunct}{\mcitedefaultseppunct}\relax
\EndOfBibitem
\bibitem[Merlin(1997)]{Merlin}
Merlin,~R. Generating coherent THz phonons with light pulses. \emph{Solid State
  Commun.} \textbf{1997}, \emph{102}, 207\relax
\mciteBstWouldAddEndPuncttrue
\mciteSetBstMidEndSepPunct{\mcitedefaultmidpunct}
{\mcitedefaultendpunct}{\mcitedefaultseppunct}\relax
\EndOfBibitem
\bibitem[Bohnen \latin{et~al.}(2001)Bohnen, Heid, and Renker]{Bohnen_mgb2001}
Bohnen,~K.-P.; Heid,~R.; Renker,~B. Phonon Dispersion and Electron-Phonon
  Coupling in ${\mathrm{MgB}}_{2}$ and ${\mathrm{AlB}}_{2}$. \emph{Phys. Rev.
  Lett.} \textbf{2001}, \emph{86}, 5771--5774\relax
\mciteBstWouldAddEndPuncttrue
\mciteSetBstMidEndSepPunct{\mcitedefaultmidpunct}
{\mcitedefaultendpunct}{\mcitedefaultseppunct}\relax
\EndOfBibitem
\bibitem[An and Pickett(2001)An, and Pickett]{An_MgB2_2001}
An,~J.~M.; Pickett,~W.~E. Superconductivity of ${\mathrm{MgB}}_{2}$: Covalent
  Bonds Driven Metallic. \emph{Phys. Rev. Lett.} \textbf{2001}, \emph{86},
  4366--4369\relax
\mciteBstWouldAddEndPuncttrue
\mciteSetBstMidEndSepPunct{\mcitedefaultmidpunct}
{\mcitedefaultendpunct}{\mcitedefaultseppunct}\relax
\EndOfBibitem
\bibitem[Perdew \latin{et~al.}(2008)Perdew, Ruzsinszky, Csonka, Vydrov,
  Scuseria, Constantin, Zhou, and Burke]{pbesol2008}
Perdew,~J.~P.; Ruzsinszky,~A.; Csonka,~G.~I.; Vydrov,~O.~A.; Scuseria,~G.~E.;
  Constantin,~L.~A.; Zhou,~X.; Burke,~K. Restoring the Density-Gradient
  Expansion for Exchange in Solids and Surfaces. \emph{Phys. Rev. Lett.}
  \textbf{2008}, \emph{100}, 136406\relax
\mciteBstWouldAddEndPuncttrue
\mciteSetBstMidEndSepPunct{\mcitedefaultmidpunct}
{\mcitedefaultendpunct}{\mcitedefaultseppunct}\relax
\EndOfBibitem
\bibitem[{van Setten} \latin{et~al.}(2018){van Setten}, Giantomassi, Bousquet,
  Verstraete, Hamann, Gonze, and Rignanese]{pseudodojo}
{van Setten},~M.; Giantomassi,~M.; Bousquet,~E.; Verstraete,~M.; Hamann,~D.;
  Gonze,~X.; Rignanese,~G.-M. The PseudoDojo: Training and grading a 85 element
  optimized norm-conserving pseudopotential table. \emph{Comput. Phys. Commun.}
  \textbf{2018}, \emph{226}, 39--54\relax
\mciteBstWouldAddEndPuncttrue
\mciteSetBstMidEndSepPunct{\mcitedefaultmidpunct}
{\mcitedefaultendpunct}{\mcitedefaultseppunct}\relax
\EndOfBibitem
\bibitem[Sangalli \latin{et~al.}(2019)Sangalli, Ferretti, Miranda, Attaccalite,
  Marri, Cannuccia, Melo, Marsili, Paleari, Marrazzo, Prandini, Bonf{\`a},
  Atambo, Affinito, Palummo, {Molina-S{\'a}nchez}, Hogan, Gr{\"u}ning, Varsano,
  and Marini]{sangalliManybody2019}
Sangalli,~D. \latin{et~al.}  Many-Body Perturbation Theory Calculations Using
  the Yambo Code. \emph{J. Phys. Condens. Matter} \textbf{2019}, \emph{31},
  325902\relax
\mciteBstWouldAddEndPuncttrue
\mciteSetBstMidEndSepPunct{\mcitedefaultmidpunct}
{\mcitedefaultendpunct}{\mcitedefaultseppunct}\relax
\EndOfBibitem
\bibitem[Bruneval and Gonze(2008)Bruneval, and Gonze]{Bruneval_GW_2008}
Bruneval,~F.; Gonze,~X. Accurate $GW$ self-energies in a plane-wave basis using
  only a few empty states: Towards large systems. \emph{Phys. Rev. B}
  \textbf{2008}, \emph{78}, 085125\relax
\mciteBstWouldAddEndPuncttrue
\mciteSetBstMidEndSepPunct{\mcitedefaultmidpunct}
{\mcitedefaultendpunct}{\mcitedefaultseppunct}\relax
\EndOfBibitem
\bibitem[Marzari \latin{et~al.}(2012)Marzari, Mostofi, Yates, Souza, and
  Vanderbilt]{MarzariMLWF2012}
Marzari,~N.; Mostofi,~A.~A.; Yates,~J.~R.; Souza,~I.; Vanderbilt,~D. Maximally
  localized Wannier functions: Theory and applications. \emph{Rev. Mod. Phys.}
  \textbf{2012}, \emph{84}, 1419--1475\relax
\mciteBstWouldAddEndPuncttrue
\mciteSetBstMidEndSepPunct{\mcitedefaultmidpunct}
{\mcitedefaultendpunct}{\mcitedefaultseppunct}\relax
\EndOfBibitem
\bibitem[Mostofi \latin{et~al.}(2014)Mostofi, Yates, Pizzi, Lee, Souza,
  Vanderbilt, and Marzari]{Mostofi_2014}
Mostofi,~A.~A.; Yates,~J.~R.; Pizzi,~G.; Lee,~Y.-S.; Souza,~I.; Vanderbilt,~D.;
  Marzari,~N. An updated version of {WANNIER90}: A tool for obtaining
  maximally-localised {Wannier} functions. \emph{Comput. Phys. Commun.}
  \textbf{2014}, \emph{185}, 2309 -- 2310\relax
\mciteBstWouldAddEndPuncttrue
\mciteSetBstMidEndSepPunct{\mcitedefaultmidpunct}
{\mcitedefaultendpunct}{\mcitedefaultseppunct}\relax
\EndOfBibitem
\end{mcitethebibliography}
\end{document}